\documentclass[aps,pra,twocolumn,a4paper,showpacs,superscriptaddress,floatfix,10pt]{revtex4}
\usepackage[pdftex]{graphicx,color}
\usepackage{dcolumn}
\usepackage{bm}

\usepackage{amsfonts}
\usepackage{amssymb}
\usepackage{amsmath}
\usepackage{amsthm}

\usepackage{subfigure}
\usepackage{rotate}

\usepackage{pdfsync}

\usepackage[pdftex]{hyperref}

\newcommand{\noshow}[1]{}

\let\oldmarginpar\marginpar
\renewcommand\marginpar[1]{\-\oldmarginpar[\raggedleft\tiny #1]%
{\raggedright\tiny #1}}

\newcommand{\ket}[1]{|#1\rangle}

\newcommand{\HH}{\mathcal{H}}

\newcommand{\up}{\uparrow}

\newcommand{\be}{\begin{equation}}
\newcommand{\ee}{\end{equation}}
\newcommand{\ba}{\begin{eqnarray}}
\newcommand{\ea}{\end{eqnarray}}

\newcommand{\bit}{\begin{itemize}}
\newcommand{\eit}{\end{itemize}}

\graphicspath{{figs/}}

\newcommand\myover[2]{\genfrac{}{}{0pt}{}{#1}{#2}}

\newcommand{\tHH}{\tilde{\mathcal{H}}}

\newcommand{\IPR}{\textrm{IPR}}

\begin{document}

\title{Localization and symmetry breaking in the quantum quasiperiodic Ising glass}

\author{A. Chandran}
\affiliation{Department of Physics, Boston University, Boston, MA 02215, USA}
\email{anushyac@bu.edu}

\author{C. R. Laumann}
\affiliation{Department of Physics, Boston University, Boston, MA 02215, USA}

\date{\today}

\begin{abstract}

Quasiperiodic modulation can prevent isolated quantum systems from equilibrating by localizing their degrees of freedom.
In this article, we show that such systems can exhibit dynamically stable long-range orders forbidden in equilibrium.
Specifically, we show that the interplay of symmetry breaking and localization in the quasiperiodic quantum Ising chain produces a \emph{quasiperiodic Ising glass} stable at all energy densities.
The glass order parameter vanishes with an essential singularity at the melting transition with no signatures in the equilibrium properties.
The zero temperature phase diagram is also surprisingly rich, consisting of paramagnetic, ferromagnetic and quasiperiodically alternating ground state phases with extended, localized and critically delocalized low energy excitations.
The system exhibits an unusual quantum Ising transition whose properties are intermediate between those of the clean and infinite randomness Ising transitions.
Many of these results follow from a geometric generalization of the Aubry-Andr\'e duality which we develop.
The quasiperiodic Ising glass may be realized in near term quantum optical experiments.

\end{abstract}

\maketitle


\section{Introduction}
\label{sec:Introduction}

\begin{figure}[b]
	\centering
	\includegraphics[width=\columnwidth]{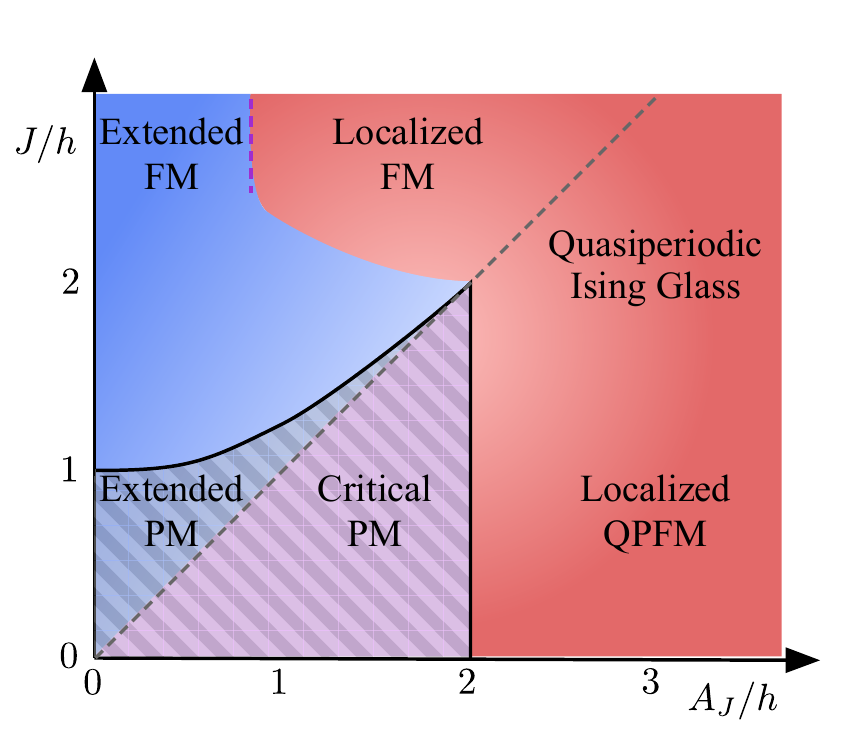}
	\caption{
	Combined symmetry breaking and localization phase diagram of the quasiperiodic Ising model at $A_h = 0$. 
	The system has quasiperiodic Ising glass excited state order at all energy densities in the red region.
	The ground state is paramagnetic (PM) in the striped region; it breaks Ising symmetry ferromagnetically (FM) above the diagonal (dashed) and with quasiperiodically alternating modulation (QPFM) below.
	The low energy excitations are localized/extended/critically delocalized in the red/blue/purple regions. 
	}
	\label{fig:phasediagram}
\end{figure}

Nearly sixty years ago, Anderson discovered that quenched disorder could localize quantum particles and thus prevent the transport necessary for equilibration in isolated systems \cite{Anderson:1958ly}.
The recent interest in the role of interactions \cite{Altshuler:1997aa,Basko:2006aa,Gornyi:2005lq,Oganesyan:2007aa,Monthus:2010vn,Vosk:2013yg,Pekker:2014aa,Pal:2010gs,Znidaric:2008aa,Bardarson:2012kl,Serbyn:2013uq,Bauer:2013rz,Serbyn:2013rt,Swingle:2013aa,Nandkishore:2014ys,Serbyn:2014aa,Iyer:2013aa,Kjall:2014aa,Laumann:2014aa,Luitz:2015fj,Chandran:2015aa,Tang:2015th,Michal:2014aa,Agarwal:2015aa,Vasseur:2015aa,Singh:2016aa,You:2016aa,Potter:2015ab} and rapid experimental developments in synthetic quantum systems \cite{Schreiber:2015aa,Kondov:2015aa,Bordia:2016aa,Choi:2016aa,Smith:2016aa,Kaufman:2016aa,Neill:2016aa,Bordia:2016ab} have led to a deeper understanding of the full range of consequences of Anderson's original observation.
The phenomenology of the localized phase is now better understood as a form of integrability with local conserved quantities \cite{Huse:2013kq,Serbyn:2013rt,Imbrie:2014jk,Ros:2015rw,chandran2015constructing,Monthus:2016aa,Rademaker:2016aa};
the dynamics of entanglement has emerged as a unifying framework for understanding thermalization \cite{Znidaric:2008aa,Bardarson:2012kl,Serbyn:2013uq};
and, the long-lived coherence of localized systems may serve as a resource for quantum information processing \cite{Serbyn:2014ek,Yao:2015aa,Bahri:2015aa}.

A particularly intriguing proposal is that localization can dynamically protect long-range order in highly excited states \emph{even} when such orders are forbidden in equilibrium  \cite{Huse:2013aa}.
The central idea may be illustrated in the 1D ferromagnetic Ising chain.
Ferromagnetic order in the ground state is usually destroyed in excited states due to the proliferation of domain walls (an argument that goes back to Peierls).
However, if quenched disorder can localize the domain walls, then the system never reaches equilibrium, and any symmetry-breaking pattern imprinted in the spin state at $t=0$ can persist for all time.
This Ising glass order clearly exists in the transverse field Ising chain in strong disorder treatments \cite{Huse:2013aa,Pekker:2014aa} and has been observed numerically in small interacting chains \cite{Kjall:2014aa}.
Localization protection has also been argued to extend to a host of more exotic orders \cite{Huse:2013aa,Chandran:2014aa,Bahri:2015aa,Potter:2016aa} and to periodically driven (Floquet) systems \cite{Khemani:2016ab,Keyserlingk:2016aa,Else:2016aa,Roy:2016aa,Po:2016aa}. 
%

Localization, however, does not require disorder, as was first recognized by Azbel \cite{Azbel:1979aa}, and Aubry and Andr\`e (AA) \cite{Aubry:1980aa} in the single particle context.
These authors discovered that sufficiently strong quasiperiodic potentials can localize a quantum particle.
Refs.~\cite{Iyer:2013aa,Michal:2014aa,Mastropietro:2015aa} extended these results to the interacting many-body case, and argued that many-body localization can persist even at high energy density.
Quasiperiodic potentials arise naturally in optical experiments using lasers with incommensurate wavevectors.
Accordingly, many experiments in such systems have now observed single-particle localization \cite{Dal-Negro:2003aa,Fallani:2007aa,Roati:2008aa,Lahini:2009aa,Modugno:2010aa,Segev:2013aa} and, more recently, have also pushed into the interacting regime and high excitation energy densities to provide evidence for the many-body localized phase \cite{Schreiber:2015aa, Bordia:2016aa, Bordia:2016ab}.

As quasiperiodic systems can show both localized and delocalized behavior already in the 1D non-interacting context, they offer a well-controlled platform to study the interplay of localization and symmetry breaking.
In this article, we study the effects of quasiperiodic modulation on the canonical quantum Ising chain. 
The most salient dynamical feature is a stable \emph{quasiperiodic Ising glass} in which all excited states exhibit Ising symmetry breaking order (red in Fig.~\ref{fig:phasediagram}).
This excited state order melts if either the ground state becomes paramagnetic or the domain wall excitations delocalize; we find both types of transition.
Remarkably, the excited state Ising glass order parameter exhibits an essential singularity at the transition, with no signatures in the ground state ordering.

In quench experiments, the quasiperiodic Ising glass phase appears in the persistence of arbitrary initial \emph{longitudinal} magnetization (i.e. in the direction flipped by the Ising symmetry) after a short transient.
This glass is accessible in current experiments in quantum optical Ising spin simulators, such as have been implemented in ion traps \cite{Islam:2011aa,Smith:2016aa} and Rydberg systems \cite{Glaetzle:2015aa,Labuhn:2016aa}. 
Experimentally, it is better to modulate the effective spin-spin interaction (as opposed to modulating the field) by quasiperiodically modulating the positions of the trapped spins, as this is the regime most favorable to finding the Ising glass. 
We have accordingly focused the detailed study in this manuscript to the coupling, rather than field, modulated regime.
Our rigorous analytic controls extends only to nearest neighbor spin-spin interactions, where the system can be fermionized, but we expect the Ising glass to persist in the presence of weak additional interactions, just as the quasiperiodically modulated many-body localized phase of bosons persists in Ref.~\cite{Iyer:2013aa}.
We discuss both potential experimental realizations and the stability to interactions further in the conclusion, Sec.~\ref{sec:Conclusions}.

From an equilibrium condensed matter perspective, the zero temperature phase diagram is interesting in its own right. 
There are paramagnetic (PM), ferromagnetic (FM) and quasiperiodically alternating ferromagnetic (QPFM) orders in the ground state.
Moreover, the low energy excitations exhibit extended, localized and critically delocalized behavior depending on the strength of the quasiperiodic modulation.
This leads to an array of possible combinations, which we have summarized in Fig.~\ref{fig:phasediagram}. 

The associated zero temperature quantum phase transitions lie in two distinct universality classes.
We find that weak quasiperiodic modulation is irrelevant at the clean Ising transition, so that the parabolic phase boundary in Fig.~\ref{fig:phasediagram} exhibits quantum critical scaling with dynamic exponent $z=1$ and extended low energy excitations. 
%
At strong modulation, we find a new quantum Ising transition separating the QPFM from the paramagnet. 
This transition exhibits dynamical critical behavior intermediate between that of the clean Ising transition and the infinite randomness transition that arises in the disordered model. 
In particular, while the correlation length diverges with $\nu = 1$, as at the clean transition, the low energy excitations undergo a transition from critically delocalized to localized, coincident with the symmetry breaking, with an apparent exponent $z = 2$. 

Our results make use of a variety of analytic and numerical techniques. 
We would like to especially flag a new relative of the celebrated Aubry-Andr\'e duality which we have discovered.
We dub this transformation `AAA triality' as it maps cyclically among three related models. 
It turns out that the `self-trial' point in the quasiperiodic Ising model sits on the phase boundary between the paramagnet and QPFM, giving us  analytic access to the unusual quantum critical properties. 
Our triality arguments explain the energy-independent wavefunction criticality on the `pure modulation', $J/h = 0$ axis in Fig.~\ref{fig:phasediagram}, which has been observed numerically before in \cite{Satija:1989aa,Cai:2014yf}. 

The non-interacting quasiperiodic models of Azbel, Aubry-Andr\'e and their generalizations have been extensively studied by mathematicians and physicists over the last thirty years for a variety of reasons \cite{Simon:1982aa,Thouless:1983aa,Wilkinson:1984aa,Sokoloff:1985wb,Schellnhuber:1987aa,Hiramoto:1989aa,Igloi:1993lr,Chaves:1997aa,Abanov:1998aa,Jitomirskaya:1999aa,Vidal:1999aa,Kraus:2012ab,Ganeshan:2015rf}.
These 1D models exhibit a single-particle localization-delocalization transition at finite modulation which mimics the metal-insulator transition in 3D disordered systems.
This is in striking contrast to the disordered Anderson model in 1D which is localized for any disorder strength \cite{Abrahams:1979ud}.
Mathematically, the models have a surprisingly rich analytic structure, exhibiting dualities, critically delocalized phases with fractal spectra and connections to higher dimensional Hofstadter-type models \cite{Aubry:1980aa,Hofstadter:1976aa}.
More generally, they offer a window into quantum localization without the epiphenomena associated with rare region effects in disordered systems.

There has been significant previous work on aperiodic and/or quasicrystalline quantum Ising chains \cite{Ceccatto:1989qm, Hermisson:1997wu,Luck:1993ad,}. 
These models have Ising couplings chosen from a finite set according to a recursive substitution rule, or by quasicrystalline projection.
There have also been several previous studies of the zero temperature properties in certain regions of the phase diagram of the incommensurately modulated Ising chain \cite{Satija:1989aa} or equivalently, in the modulated p-wave superconductor \cite{DeGottardi:2013th,Cai:2013aa,Cai:2014yf}.
The AAA triality we introduce provides an analytic framework for explaining numerical observations in these works.

The organization of the paper is as follows. We begin in Sec.~\ref{sec:GenProperties} with a precise definition of the Ising model, a review of its fermionization and various salient facts about quasiperiodic modulation in chains. 
While much of Sec.~\ref{sec:GenProperties} is review, our geometric interpretation of the AA duality in two dimensions may provide an alternative perspective for many readers.
In Sec.~\ref{sec:GroundStateSymmetryBreaking}, we derive the ground state symmetry breaking phase diagram.
We turn to the localization properties of the low energy excitations in Sec.~\ref{sec:localization_of_excitations}.
With these basic properties in hand, we discuss the zero temperature quantum critical behavior in Sec.~\ref{sec:GroundStateTransitions}.
We investigate the properties of the excited state Ising glass order in Sec.~\ref{sec:excited_glass_order} and its melting transition in Sec.~\ref{sub:excited_state_transition}. 
We conclude with a discussion of the role of interactions, possible experimental realizations and other open questions.



\section{General properties of the model}
\label{sec:GenProperties}

\begin{figure}[tbp]
	\begin{center}
	\includegraphics[width=\columnwidth]{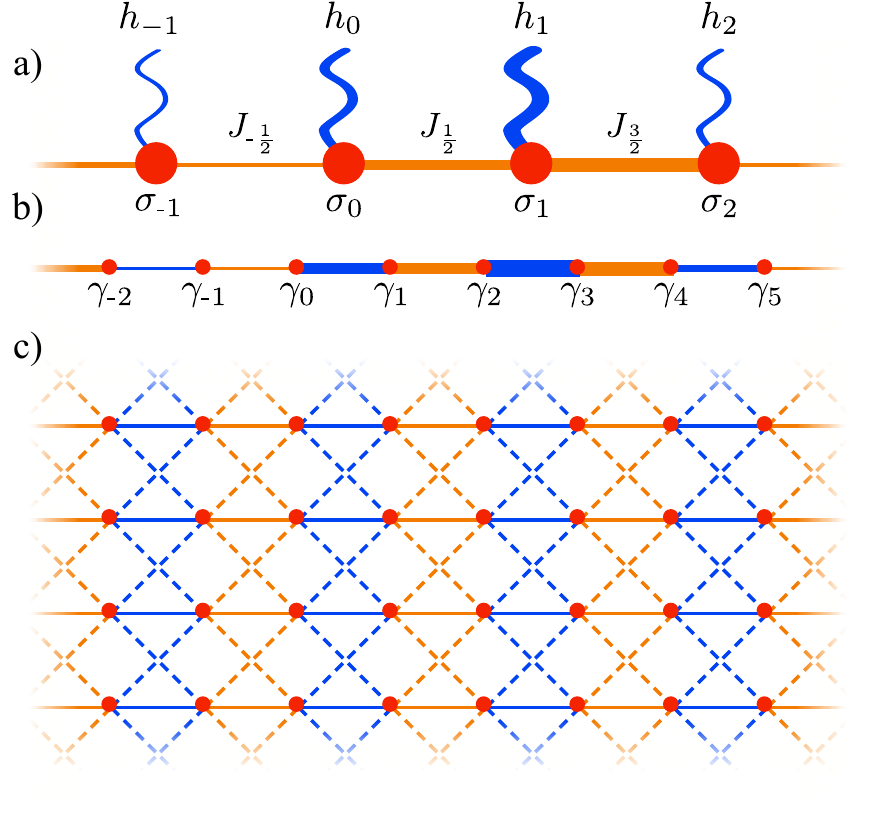}
	\caption{%
	(a) The transverse field Ising chain with spatially varying fields $h_i$ (blue) and bonds $J_{i+\frac{1}{2}}$ (orange).
	(b) The Jordan-Wigner transformation maps the spins $\sigma_i$ to Majorana fermions $\gamma_{2i}, \gamma_{2i+1}$ arranged in a `hopping' chain.
	(c) The two-dimensional hopping model Eq.~\eqref{Eq:2DHofstadterHam} obtained by treating the phase $\phi_h$ as quasi-momentum $k_y$ and inverting the Fourier transformation. Each unit cell (of two sites) is pierced by uniform flux $Q$. 
	The solid/dashed blue bonds have hopping strength $ h$/$A_h$ while the solid/dashed orange bonds correspond to $ J$/$A_J$. 
	}
	\label{fig:TFIMcouplings}
	\end{center}
\end{figure}

The Hamiltonian of the one-dimensional quasi-periodic transverse field Ising model (TFIM) is:
\begin{align}
	\HH &= -\frac{1}{2} \sum_{j} J_{j+1/2} \sigma_j^x \sigma_{j+1}^x + h_j \sigma_j^z \label{Eq:hidef} \\
    h_j &=  h + A_h \cos(Q j + \phi + \Delta) \label{Eq:HcouplingDef} \\
	J_{j+1/2} &=  J + A_J \cos(Q (j+1/2) + \phi) \label{Eq:JcouplingDef} 
\end{align}
The model is illustrated in Fig.~\ref{fig:TFIMcouplings}(a). 
Here, $\sigma_j^\alpha$ are the Pauli matrices with $j \in \mathbf{Z}$ running over the sites in the chain and $\alpha = x,y,z$, $Q$ is the wavevector of the modulation in units where the lattice spacing is $a=1$, and the phases $\phi$ and $\phi+\Delta$ shift the positions of the maxima of the couplings relative to the underlying lattice.
The wavevector $Q$ is commensurate with the underlying lattice if $Q/2\pi = p/q$ is rational, and is incommensurate otherwise.
We choose the wavevector of the modulation of the Ising coupling and the transverse field to be the same for simplicity and as this is natural if the modulation arises from the same physical source (eg. an incommensurate laser potential).

\emph{Global symmetries:}
The quasi-periodic TFIM has several global symmetries.
The eponymous Ising symmetry is given by $G = \prod_{i} \sigma_i^z$ -- this is the symmetry which breaks spontaneously in the $T=0$ ferromagnetic and localized Ising glass phases. 
The Hamiltonian $\HH$ is also symmetric under complex conjugation $K$, which is anti-unitary. 
Finally, for special values of the modulation phases $\phi$ and $\Delta$ the model is symmetric under reflections across site $k$, $j \to k - j$ or bond $k+1/2$, $j \to k + 1/2 - j$. 
For example, at $\phi=0$ and $\Delta = 0$, $\HH$ is symmetric under $j \to - j$.

\emph{Ising duality:}
Under the duality transformation
\begin{align}
\sigma_i^{x}\sigma_{i+1}^x &= \tau_{i+1/2}^z \\
\sigma_i^z &= \tau_{i-1/2}^x \tau_{i+1/2}^x 
\end{align}
$\HH$ maps onto another incommensurate TFIM $\HH'$ with the role of the field and bond couplings interchanged (and different boundary conditions).
%
Formally, the duality maps
\begin{align*}
	h' &= J  & A_h' &= A_J \\
	J' &= h  & A_J' &= A_h \\
	\phi' &= \phi + \Delta & \Delta' &= - \Delta
\end{align*}
The duality swaps  paramagnetic and ferromagnetic phases but leaves the dynamical nature of the bulk single-particle excitations (spectrum and wavefunction localization) invariant. 
This can be seen most easily from the fermionization (see below) of $\HH$ and its dual $\HH'$, whose non-interacting Hamiltonians agree precisely up to a translation by half a unit cell, so that their single fermion modes are identical up to this half translation.

\emph{Fermionization:}
The TFIM has a well-known fermionic representation which we review and extend to the quasiperiodic case here.
The Jordan-Wigner transformation introduces a pair of Majorana fermion operators for each spin-1/2:
\begin{align}
\gamma_{2i} &= \left(\prod_{j<i} \sigma_j^z\right) \sigma_i^x, &
\gamma_{2i+1} &= \left(\prod_{j<i} \sigma_j^z\right) \sigma_i^y
\label{Eq:JordanWigner}
\end{align}
The $\gamma$ operators are Hermitian and satisfy the canonical anti-commutation relation $\{\gamma_i, \gamma_j\} = 2\delta_{ij}$.
The transformation maps the TFIM to a quadratic Majorana chain with Hamiltonian:
\begin{align}
\label{Eq:MajoranaH}
	\HH = \frac{i}{2} \sum_{j} \left[ J_{j+1/2} \gamma_{2j+1} \gamma_{2j+2} + h_j \gamma_{2j}\gamma_{2j+1} \right] 
\end{align}
See Fig.~\ref{fig:TFIMcouplings}(b).

The dynamical and symmetry-breaking properties of the TFIM Hamiltonian $\HH$ follow from the properties of the single-particle Hamiltonian $H$ defined by
\begin{align}
	\HH &= \frac{1}{4} \sum_{ij} \gamma_i H_{ij} \gamma_j .\label{Eq:MajoranaHam}
\end{align}
Hermiticity of $\HH$ requires that $H_{ij}$ is an imaginary anti-symmetric $2L\times 2L$ matrix where $L$ is the number of spins. 
Thus, the eigenvalues of $H$ come in $\pm e$ pairs corresponding to complex conjugate eigenmodes, $\psi$ and $\overline{\psi}$. 
Labeling the $L$ positive energy eigenmodes by the index $\alpha$, we can diagonalize $\HH$ into the familiar form
\begin{align}
	\HH &= \sum_\alpha e_\alpha c^\dagger_\alpha c_\alpha 
\end{align}
where
\begin{align}
    c^\dagger_\alpha &= \sum_j \psi^\alpha_j \gamma_j &
    c_\alpha &= \sum_j \overline{\psi^\alpha_j} \gamma_j
\end{align}
The complex fermion operators $c, c^\dagger$ satisfy the usual anti-commutation relations
\begin{align*}
	\{ c^\dagger_\alpha, c^\dagger_\beta \} = \{ c_\alpha, c_\beta \} = 0, \qquad
    \{ c^\dagger_\alpha, c_\beta \} = \delta_{\alpha \beta}
\end{align*}

In the ground state, the paramagnetic phase corresponds to the topologically trivial phase of the Majorana chain, while the ferromagnetic phase maps to the topologically non-trivial phase.
The simplest way to detect the topological phase of the Majorana chain is with open boundary conditions, in which case the topologically non-trivial phase possesses a pair of zero energy Majorana modes localized at the boundaries of the chain \cite{Kitaev:2001aa}.
The many-body ground state space is accordingly doubly degenerate, as the fermionic mode defined by the two zero energy Majorana operators can be occupied or unoccupied at zero cost.
We use this approach to extract the ground state phase diagram of the quasiperiodic model in Sec.~\ref{sec:GroundStateSymmetryBreaking}.

The symmetries of Eq.~\eqref{Eq:hidef} appear in the fermionic language as follows. 
The global Ising symmetry operator $G = \prod_j \sigma^z_j$ maps to the fermionic parity operator $G = \prod_j (-i \gamma_{2j} \gamma_{2j+1})$, while the symmetry under complex conjugation forces $H$ to be bi-partite.
The action of the Ising duality on the Majorana chain shifts all the site labels by a half: $j \to j - 1/2$ as mentioned above.

As all eigenstates of $\mathcal{H}$ correspond to Slater determinant states of the fermions $\gamma$, they all satisfy Wick's theorem. This allows evaluation of the spin-spin correlation function,
\begin{align}
	\langle \psi| \sigma^x_i \sigma^x_j |\psi \rangle &= -i \langle \psi | \gamma_{2i+1}\prod_{k=i+1}^{j-1}(-i \gamma_{2k} \gamma_{2k+1}) \gamma_{2j}| \psi \rangle 
\end{align}
as a Pfaffian of the fermionic Green function
\begin{align}
 	G^\psi_{ij} &= \langle \psi | \gamma_i \gamma_j | \psi \rangle  - \delta_{ij}
\end{align} 
restricted to the diagonal block from $2i+1$ to $2j$. 
While this representation is not easy to use analytically, it allows straightforward numerical computations of the exact correlation functions in  large systems (eg. up to $L=1000$ in this work). 
Evaluating these correlators at large separation $|i - j|$ allows us to numerically extract the magnetization as
\begin{align}
	\langle\psi| \sigma^x_i \sigma^x_j |\psi \rangle &\xrightarrow{|i-j|\to\infty}M^\psi_i M^\psi_j
\end{align}
where $M^\psi_i$ is the magnetization of spin $i$ in state $\ket{\psi}$.

\emph{The 2D model:}
The incommensurate TFIM has a second useful representation in terms of a 2D model of non-interacting complex fermions.
To derive this, consider first a complex fermionic model with the same single particle Hamiltonian as Eq.~\eqref{Eq:MajoranaHam} \footnote{This is often referred to as a `doubled' model of the Majorana fermions since it corresponds to two non-interacting copies. We denote the Hamiltonians of `doubled' fermion models by $\tHH$}:
\begin{align}
&\tHH_{1D}(\phi) = \sum_{ij} d^\dagger_i H_{ij} d_j \nonumber\\
&\,\,=  i\sum_{j} \left[ J_{j+1/2} d^\dagger_{2j+1} d_{2j+2} + h_j d_{2j}^\dagger d_{2j+1}\right] + h.c.  \label{Eq:1DComplexFermionModel}
\end{align}
Above, $d_i$ destroys a fermion at site $i$ and $d_i, d_j^\dagger$ satisfy the usual complex anti-commutation relations.
On varying the phase $\phi$ between $[0,2\pi)$, we generate a family of distinct 1D Hamiltonians.
Treating $\phi$ as the momentum along an extra dimension $y$ and inverting the Fourier transformation, we obtain a 2D tight binding model:  
\begin{align}
\tHH_{2D} &=  i\sum_{j,k} \left( J d^{\dagger}_{2j+1,k} d_{2j+2,k} +  h d^{\dagger}_{2j,k}d_{2j+1,k} \right. \nonumber\\
 &+ \frac{A_J}{2} e^{iQ(j + 1/2)}d^{\dagger}_{2j+1,k} d_{2j+2,k+1}   \nonumber \\
 &+\frac{A_J}{2}e^{-iQ(j  + 1/2)}d^{\dagger}_{2j+1,k}d_{2j+2,k-1}  \nonumber \\ 
 & + \frac{A_h}{2}e^{i(Qj + \Delta)} d^{\dagger}_{2j,k} d_{2j+1,k+1}  \nonumber\\
 &\left. + \frac{A_h}{2}e^{-i(Qj + \Delta)}d^{\dagger}_{2j,k}d_{2j+1,k-1}  \right) + h.c.
 \label{Eq:2DHofstadterHam}
\end{align}
whose spectrum at fixed y-momentum $\phi$ reproduces the spectrum of the 1D model $\tilde{H}_{1D}(\phi)$, see Fig.~\ref{fig:TFIMcouplings}(c).
The Hamiltonian $\tilde{H}_{2D}$ describes a translation invariant hopping model with uniform flux $Q$ piercing each two-site unit cell in Landau gauge. 
The flux associated with hopping cycles within a unit cell depends on $\Delta$. 
There are no vertical hops in the 2D model (ie. from $(j,k)$ to $(j,k\pm1)$) in Fig.~\ref{fig:TFIMcouplings}(c) because the 1D Hamiltonian in Eq.~\eqref{Eq:1DComplexFermionModel} is off-diagonal.

The localization properties of the excitations of the 1D TFIM map onto those of the $\tilde{H}_{2D}$ in the Landau gauge at fixed y-momentum $\phi$. 
By construction, if the eigenstates of the 2D model are delocalized in the $x$-direction, then the eigenstates of the corresponding 1D model are extended, while if the states of the 2D model are localized in the $x$-direction, then the 1D model is localized.
As we will see, the 2D picture is a surprisingly useful geometric aid for identifying localized, extended and critical phases of the excitations.

\emph{Aubry-Andr\'e model:}
We will need several properties of the original Aubry-Andr\'e hopping chain in the analysis of the quasiperiodic TFIM.
The AA model has the following Hamiltonian \cite{Aubry:1980aa},
\begin{align}
\label{Eq:HamAA}
\HH_{AA} = \sum_j  -t(d_j^\dagger d_{j+1} + h.c.) - 2V \cos(Qj+\phi_V) d_j^\dagger d_j.
\end{align}
For $V>t$, the single particle states are localized at all energies while for $V<t$ they are extended at all energies. 
At the critical point, $V=t$, the states exhibit multifractal properties.
Further, the localization length diverges at the critical point with exponent $\nu = 1$. More precisely,
\begin{align}
	\xi = 1/\log|V/t|\sim |V/t-1|^{-1}
\end{align}

Many features of the phase diagram follow Aubry-Andr\'e duality. 
This duality corresponds to a $\pi/2$ rotational symmetry of the associated 2D model. 
The 2D model is that of a particle hopping on an anisotropic square lattice with flux $Q$ per plaquette and hopping strength $t$ and $V$ in the $x$ and $y$ directions, respectively; at $t=V$, this is the Hofstadter model, whose fractal character is well known \cite{Hofstadter:1976aa}.
The $\pi/2$ rotation swaps $t$ and $V$ and accordingly swaps the localized and extended phases of the original 1D model; clearly, $t=V$ is self-dual.
Moreover, the rotation ensures the localization properties are energy independent for any $t$ and $V$, as we will see in more detail in Sec.~\ref{sec:localization_of_excitations} from an analysis of the characteristic polynomial.

\emph{Lack of Aubry-Andr\'e duality:}
The incommensurate TFIM clearly lacks the Aubry-Andr\'e rotational symmetry, as Fig.~\ref{fig:TFIMcouplings}(c) does not map onto itself (up to swapping couplings) under rotation by $\pi/2$.
While one can embed the 2D model into a larger class of 2D models which have the requisite vertical bonds, these would in general need to be staggered in the $y$-direction and thus they do not correspond to 1D incommensurate chains.
The $\pi/2$ rotation in the larger model space therefore does not define a duality map on the incommensurate TFIM.

This is both blessing and curse: the incommensurate TFIM exhibits rich phenomena not allowed by AA duality, such as energy and $Q$-dependent mobility edges, but it is correspondingly harder to rigorously analyze.
In Sec.~\ref{sec:localization_of_excitations}, we will see that various limits of the  quasiperiodic model have higher rotational symmetry when viewed in 2D. These will give us analytic control of the phase diagram in those limits.

\emph{Parameter regime we study:}
Except for special lines in coupling space, even the ground state phase diagram has not been studied. 
The full dynamical phase diagram lives in an eight-dimensional space, parameterized by $h$, $J$, $A_h$, $A_J$, $Q$, $\phi_h$, $\Delta$ and the excitation energy $e$; after normalizing the units of energy, there are still 7 dimensionless parameters. 
Where simple enough, we will provide general expressions that apply to the entire phase diagram.
However, as seven dimensional phase diagrams are unwieldy, we mostly focus on the manifold defined by $A_h=0$, which turns out to already be rather interesting.
In units where $ h=1$, the relevant parameters controlling the phase diagram are then $ J$, $A_J$, $Q$, $\phi$ and the excitation energy $e$ (note that $\Delta$ has no effect when $A_h=0$).
Most of the analytic features we derive hold for any incommensurate $Q$ and $\phi$, but we focus our numerical results (notably for spectra) on $Q/2\pi = (\sqrt{5}+1)/2$, the golden mean.
By Ising duality, our analysis also produce the phase diagram at $A_J=0$.

\section{Ground state symmetry breaking}
\label{sec:GroundStateSymmetryBreaking}

\begin{figure*}[t]
	\centering
	\includegraphics[width=\textwidth]{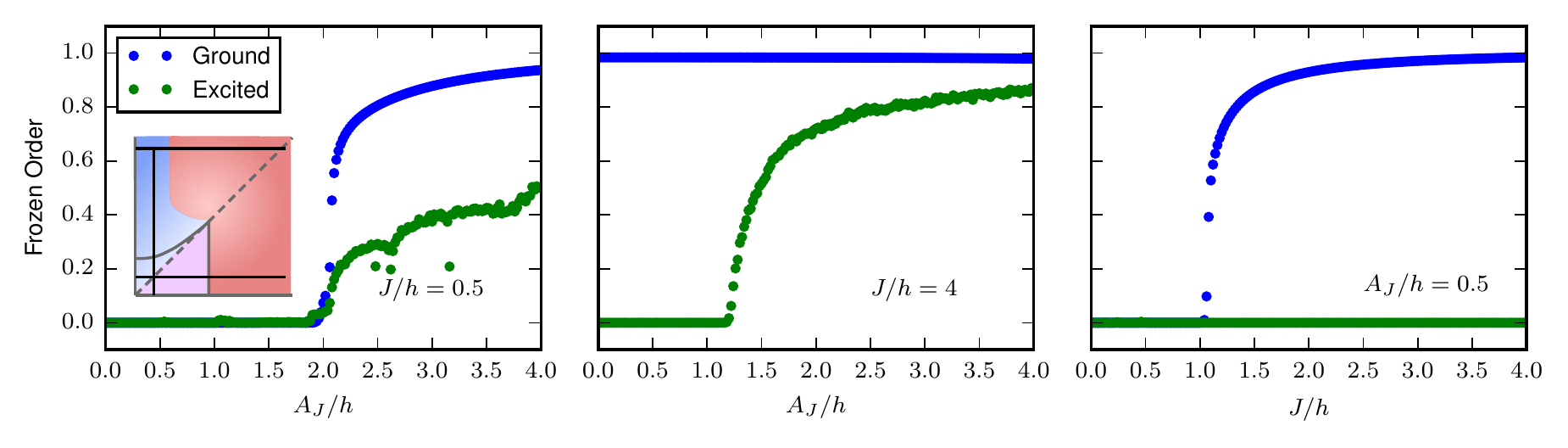}
	\caption{The Ising glass order parameter along three representative cuts in the phase diagram (inset).
	The definitions of the ground and excited state order parameters are given by Eqs.~\eqref{eq:gsorderparameter} and \eqref{eq:excitedorderparameter}, respectively. 
	The spatial averaging window is of size $w = 5$ and we sample $50$ eigenstates in the excited state average.
	In the first panel, both the ground and infinite temperature order turn on at the same coupling as the excitations localize across the transition.
	In the second panel, the ground state order is unchanged even as the excited state order turns on due to the localization of the domain walls.
	In the third panel, the ground state order turns on but the excitations remain extended so that the excited state order remains zero.}
	\label{fig:frozenmag}
\end{figure*}

The quasiperiodic Ising model has three ground state phases: a paramagnetic (PM)phase, a ferromagnetic (FM) phase and a quasiperiodically alternating ferromagnetic (QPFM) phase.
The two latter phases spontaneously break the Ising symmetry and the spin-spin correlation function $\langle\sigma^x_i \sigma^x_j\rangle$ has long-range order.
In the simple FM, all of the spins magnetize in the same direction (although the magnetization is not spatially uniform), while in the QPFM, the spatial mean magnetization is zero due to the presence of antiferromagnetic links where $J_j < 0$
We refer to both Ising ordered phases as ``ferromagnetic'' as the ordering even in the QPFM is unfrustrated (ie. gauge equivalent to a $J_j>0$ model).

The clean model ($A_J=A_h=0$) spontaneously breaks the Ising symmetry for $|J| > |h|$. 
On general grounds, the corresponding gapped ferro- and paramagnetic phases should persist in the presence of small incommensurate modulation $A_h, A_J$. 
To find the phase boundaries in general, we recall that the Ising symmetry breaking phase corresponds to the topological phase of the fermionic representation, Eq.~\eqref{Eq:MajoranaH}, which famously hosts zero-energy Majorana modes, $\Gamma_{L}$ and $\Gamma_{R}$, bound to the left and right ends of an open chain \cite{Kitaev:2001aa}.
To detect ground state symmetry breaking, we look for normalizable boundary modes.

For simplicity, we focus on the left edge of a semi-infinite chain. 
As the Hamiltonian Eq.~\eqref{Eq:MajoranaH} is bipartite, only connecting the even and odd sublattices, the zero mode localized at the left edge must have the following form:
\begin{align}
\Gamma_L = \sum_{j=0}^{\infty} \alpha_j \gamma_{2j}
\end{align}
(The right mode would be localized on the odd sublattice.)
Substituting in the eigenvalue equation $[\HH,\Gamma_L] =0$, we obtain:
\begin{align}
\alpha_{j+1} &= \frac{h_j}{J_{j+1/2}} \alpha_j 
\end{align}
Taking a logarithm, we obtain
\begin{align}
S & \equiv \log(\alpha_j/\alpha_0) \nonumber\\
&= \sum_{i=0}^{j-1} \log|h_i| - \sum_{i=0}^{j-1} \log|J_{i+1/2}|\label{Eq:alphaj}
\end{align}
In order for the zero mode to be normalizable, $S_j$ must decrease sufficiently rapidly with $j$ at large distances.

In the absence of modulation, $S = j \log |h|/|J|$ clearly grows linearly with $j$; the sign of $\log |h| / |J|$ thus determines the normalizability of $\Gamma_L$. 
Modulation at irrational wavevector $Q$, causes $S_j$ to fluctuate, but on distances long compared to $Q^{-1}$, we expect $S_j$ to still have a linear trend as the sum averages over the modulation.
This trend can be extracted formally by averaging over the phase,
\begin{align}
S &\sim j I \\
I & =\int_0^{2\pi} \frac{d\theta}{2\pi} \log\left|\frac{ h + A_h \cos(\theta)} { J + A_J \cos(\theta)}\right|
\label{Eq:IIntegral}
\end{align}
The sign of $I$ determines if the zero modes are normalizable ($I < 0$) or not ($I>0$). 
The Ising phase boundary is at $I=0$, which is shown by the black curve on the phase diagram at $A_h = 0$ in Fig.~\ref{fig:phasediagram}.
We evaluate $I$ in Appendix~\ref{app:ground_state_phase_boundary}.
The contour lies at,
\begin{align}
	\label{eq:gsphaseboundary}
	J/h &= 1 + (A_J/2h)^2 & \textrm{for } J > A_J \nonumber \\
	A_J/h &= 2 & \textrm{for }J < A_J
\end{align}

In Appendix~\ref{app:phaseboundaries}, we present an alternative derivation of the phase boundaries by approaching the incommensurate $Q$ limit through a series of longer and longer commensurate wavenumbers $2\pi p/q \to Q$. 
This provides a more rigorous treatment of the role of incommensuration. 
The two derivations are in perfect agreement.

Numerical evaluation of long-range order in the ground state agrees with the phase boundary determined from the above analysis, Eq.~\eqref{eq:gsphaseboundary}. 
In Fig.~\ref{fig:frozenmag}, the three panels present the average magnitude of the magnetization squared on representative cuts through the phase diagram, calculated at $Q/2\pi = (1+\sqrt{5})/2$ for open chains of length $L=1000$. 
More precisely, we detect the ground state order through the correlator,
\begin{align}
	\label{eq:gsorderparameter}
	\mathcal{O} = [|\langle 0 | \sigma^x_{L/4+j} \sigma^x_{3L/4 + j} | 0\rangle|]_j 
\end{align}
which is non-zero in both the FM ($A_J < J$) and QPFM ($A_J > J$) phase.
The square brackets $[\cdot]_j$ indicate averaging over a small spatial window ($j=-w,\cdots, w$) to smooth out the quasiperiodic modulation of the magnitude. 
In the thermodynamic limit $L\to\infty$, we expect
\begin{align}
	\mathcal{O} &\to [|M|]^2 + O\left(\frac{1}{Q w}\right)
\end{align}
where $[|M|]$ is the average of the absolute value of the site magnetization and $w$ is the width of the spatial averaging window.

We end with a few comments. 
First, the general derivation reproduces the critical point of the clean $A_J = A_h = 0$ model. 
Second, Ising duality immediately implies that the phase diagram in the $A_h/J$, $h/J$ plane at $A_J=0$ looks identical to that in Fig.~\ref{fig:phasediagram} after swapping the ferromagnetic and paramagnetic phases. 
Third, whether the ground state breaks Ising symmetry or not is independent of $Q$, so long as $Q$ is incommensurate, as only the average $\log |h|/|J|$ over a period matters to the long-distance behavior. 
Fourth, the fluctuations of $S$ relative to its linear trend are $O(1)$ in distance $j$. This is significantly less than the $O(\sqrt{j})$ fluctuations that develop in the case of usual disorder with independent random couplings.
Finally, the cusp at $A_J = J = 2$ is real and indicates that the transition on the parabolic boundary between the FM and PM and on the vertical boundary between the QPFM and PM have different character. 
We will return to this in greater detail in Sec.~\ref{sec:GroundStateTransitions}.


\section{Localization of excitations}
\label{sec:localization_of_excitations}

The dynamical phase diagram of the quasiperiodic TFIM follows from the properties of the fermionic excitations described by Eq.~\eqref{Eq:MajoranaH}. 
The main feature of interest for the stability of excited state order is the spatial extent of the single particle wavefunctions.
That is, whether they are extended, critically delocalized or localized. 
In general, at any given point in the phase diagram, these properties are energy dependent and the system may exhibit mobility edges. These can be studied numerically quite effectively. 
However, there are many special lines in the coupling space with enhanced symmetry which provides analytic control and energy independence. 
On the $A_h = 0$ plane represented in Fig.~\ref{fig:phasediagram}, both the axes, $A_J = 0$ and $J = 0$, and the large coupling limit, $J\to\infty$, have such enhanced symmetry. 
These limits will be sufficient to asymptotically characterize all of the features in Fig.~\ref{fig:phasediagram}.
We study these limits in the subsections below before turning to numerics to support the bulk of the phase diagram.

The simplest diagnostic of localization is the inverse participation ratio defined for a given eigenstate $\alpha$ as
\begin{align}
	\IPR &= \sum_{j} |\psi_j^\alpha|^4.
\end{align}
In finite size studies, the scaling of the IPR in a given energy window, $\IPR \sim 1/L^\gamma$, detects the dynamical phase. 
In the extended phase, $\gamma = 1$; in the localized, $\gamma = 0$; and, in the critical, $0<\gamma<1$. 
Formally, these phases correspond to spectra that are absolutely continuous, pure point and singular continuous (fractal) respectively.
We will use both diagnostics in the following analysis.

\subsection{Clean Limit $A_J \to 0$} 
\label{sub:clean_limit}

In the absence of the incommensurate modulation, the model reduces to the usual nearest neighbor Ising model. 
It has extended excitations for all parameter values ($J/h$) and at all energies. 

The usual Ising critical point at $J/h = 1$ has gapless extended excitations at all energies. 
As we argue in more detail in Sec.~\ref{sec:GroundStateTransitions}, the parabolic ground state phase boundary extending from the clean critical point ($J > A_J$, Eq.~\eqref{eq:gsphaseboundary}) lies in the same universality class and thus we expect the \emph{low} energy excitations to remain extended all along this boundary.
However, mobility edges are allowed at higher energy.
These features are visible in the numerical data shown in Fig.~\ref{fig:cuts-enipr}(a,b).


\subsection{Large $J \gg A_J, h$} 
\label{sub:large_Jlimit}

In this regime, the ground state is very close to the ideal ferromagnet with all the spins pointing in the $+x$ or $-x$ direction. 
The excitations are domain walls, 
\begin{align}
\ket{j} = | \cdots \leftarrow_{j-1} \leftarrow_j \vert \rightarrow_{j+1} \rightarrow_{j+2} \cdots \rangle
\end{align}
Up to a constant, the Hamiltonian for a domain wall is:
\begin{align}
H_{1DW} = -\frac{1}{2} \sum_{j} \left[ h |j \rangle \langle j+1 | + h.c. - 2 J_{j+1/2} |j \rangle \langle j| \right]
\end{align}
This is simply the AA model, Eq.~\eqref{Eq:HamAA}, with $t = h$ and $V = A_J$. Thus, the domain walls are extended at all energies for $A_J/h<1$ and localized at all energies for $A_J/h > 1$ formally at $J \to \infty$. 
This explains the vertical asymptote of the extended to localized phase boundary at large $J$ in Fig.~\ref{fig:phasediagram}.

Note one can also see the emergence of the AA model in this limit by considering the associated 2D model. In the large $J \gg A_J, h$ limit, one pairs the fermionic sites across the horizontal $J$ links and obtains an effective square lattice with horizontal links of strength $h$ and vertical links of strength $A_J$.


\subsection{The Corner $J, A_J \ll h$} 
\label{sub:the_corner}

This regime is in the bottom left corner in Fig.~\ref{fig:phasediagram}. 
The ground state is paramagnetic and is very close to the product state $\ket{\up\up\up\cdots\up}$. 
The excitations are spin flips. Analogously to Sec.~\ref{sub:large_Jlimit}, the effective Hamiltonian for the spin flips is given by,
\begin{align}
\label{Eq:GenAAModel}
H_{1F} &= -\frac{1}{2} \sum_{j} \left(J + A_J \cos(Q(j+1/2)+\phi)\right) |j\rangle \langle j+1| \nonumber \\
&+ h.c. 
\end{align}
up to a constant.
This off-diagonal Aubry-Andre model admits the AA duality and accordingly has energy independent localization properties. 
The model has been previously studied by Thouless and co-workers \cite{Thouless:1983aa, Han:1994cs}, who found two dynamical phases: extended for $A_J <  J$, and critical for $A_J >  J$. 
Asymptotically, this coincides with the diagonal dashed line near the origin of Fig.~\ref{fig:phasediagram}.

As an aside, the critical phase for $A_J > J$ is a special feature of the $A_h=0$ plane.
At small $A_h$ and $\Delta\neq 0$, the critical phase is destroyed and the states for $A_J > J$ are fully localized \cite{Liu:2015aa}.

\subsection{Large $A_J \gg J,h$} 
\label{sub:large_amplitude}

The 2D model in the $A_J \to \infty$ limit reduces to a collection of decoupled vertical zig-zag wires (dashed orange in Fig.~\ref{fig:TFIMcouplings}), two per original spin.
Without $h$, this is clearly localized for the 1D model, as it decouples at every bond.

To determine the stability of the localization to small $J$ and $h$, note that the pair of wires at position $j$ have dispersion 
\begin{align}
	e_j(\phi) &= \pm |J + A_J \cos(\phi + Q(j+1/2))|,
\end{align}
where $\phi$ is the $y$-momentum \footnote{Technically, the $\pm$ energy eigenstates at momentum $\phi$ are delocalized on both wires at position $j$.}. 
The energy dispersions are shifted by $Q$ between adjacent wire pairs.
Thus, for incommensurate $Q \sim O(1)$, neighboring wire states at fixed momentum $\phi$ are typically non-degenerate, with an energy splitting of order $A_J$.
To leading order, this implies that $h$ is typically an off-resonant perturbation and localization in the $x$ direction persists. 
This argument can be generalized to higher order using Diophantine properties of $Q$ and closely mirrors arguments which can be made in the anisotropic Hofstadter problem.

Thus, we expect that all states are localized at large $A_J \gg J, h$ to the far right of Fig.~\ref{fig:phasediagram}. We note that the 2D model does not have enhanced symmetry in this limit and thus the analysis does not provide as complete control as it does in the Hofstadter model.

\subsection{Pure modulation $J=0$ and AAA Triality} 
\label{sub:pure_modulation_aaa_triality}

The analysis on the $J=0$ axis of Fig.~\ref{fig:phasediagram} is significantly more involved than the previous limits, as it turns out that the model exhibits a hidden symmetry that allows us to define a `AAA tri-ality' operation in analog with AA duality. 
The existence of this triality implies that the localization properties of the wavefunctions are independent of energy, as in the AA model.
It also enables an exact evaluation of certain properties of the secular equation, from which we find that the spectrum is critical for $A_J/h < 2$ and localized for $A_J/h > 2$. 
The self-`tri-al' point lies at $A_J/h = 2$.
We note that this explains various numerical observations made on this axis in previous studies of the p-wave superconducting chain \cite{Satija:1989aa,DeGottardi:2013th,Cai:2013aa,Cai:2014yf}.

\begin{figure}[tb]
	\centering
	\includegraphics[width=\columnwidth]{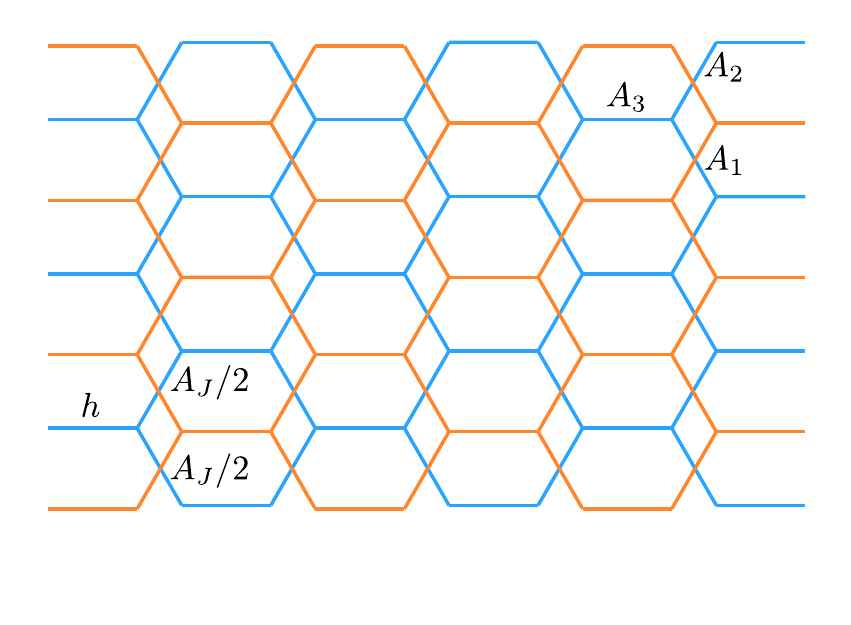}
	\caption{The 2D model associated to the incommensurate TFIM at $A_h = 0, J=0$ decouples into two interpenetrating layers (blue and orange) of honeycomb lattice. Each hexagonal plaquette is pierced by flux $2Q$. We have adjusted the geometric embedding of the lattice points relative to Fig.~\ref{fig:TFIMcouplings} in order to emphasize the symmetry associated with rotation by $2\pi/3$. 
	All parallel bonds carry the same coupling magnitude (though the hopping phase depends on the choice of gauge). 
	The incommensurate TFIM at $J=0$ embeds into a larger AAA family of models with couplings $A_1 = A_J/2, A_2 = A_J/2, A_3=h$ on bonds as shown. }
	\label{fig:honeycomblayers}
\end{figure}

The AAA triality is best understood as a geometric transformation of the associated 2D model.
At $J=0$, this model decouples into two interpenetrating honeycomb lattice `layers' each pierced by flux $2Q$ per hexagonal plaquette, see Fig.~\ref{fig:honeycomblayers}. 
At $A_J/h = 2$, since the layers are decoupled, the system is symmetric under rotation by $2\pi/3$ independently in each layer, about any site. 
These are physical rotations in the 2D model, which need to include gauge transformations for $\tilde{H}_{2D}$ to return to Landau gauge. 

In the usual AA model, $\pi/2$ rotation in the associated 2D model leads to another model with couplings $V$ and $t$ swapped.
Rotating both layers by $2\pi/3$ in the honeycomb model rotates the couplings associated with the three bond angles into one another. 
This suggests the utility of generalizing the 2D model to have three independent couplings, $A_1$, $A_2$ and $A_3$ corresponding to the three different bonds.
In Landau gauge, this corresponds to a 1D incommensurate model with Hamiltonian,
\begin{align}
\label{eq:HamiltonianAAA}
	\tHH_{AAA}(\phi) = \sum &\left( i A_3 d^\dagger_{2j} d_{2j+1} \right. \nonumber \\
	&+ i A_1 e^{i (Q(j+1/2) + \phi)} d^\dagger_{2j+1} d_{2j+2} \nonumber\\
	& \left. + i A_2 e^{-i (Q(j+1/2) + \phi)} d^\dagger_{2j+1} d_{2j+2}\right) + h.c.
\end{align}
The original 1D TFIM at $J=0$ corresponds to $A_3 = h$, $A_1 = A_2 = A_J/2$. 
In the extended AAA family of models, the $2\pi/3$ rotation maps $\tHH_{AAA}$ to $\tHH_{AAA}'$ with cyclically permuted couplings.
This is the AAA triality.

With this triality in hand, it is possible to analyze the wavefunction properties of the AAA model in considerable detail. 
We relegate this analysis to the Appendices, as it entails a considerable calculational detour.
The upshot is that the 1D incommensurate TFIM has a critical to localized transition, at all excitation energies, at the self-trial point, $A_J = 2h$. 

\subsection{Numerical support away from the limits} 
\label{sub:numerical_support_away_from_the_axes}

Away from the special lines discussed in the previous subsections, the localization properties of the wavefunctions are both $Q$ and energy dependent. 
We rely on numerics to confirm the features summarized in Fig.~\ref{fig:phasediagram}.
We have extensively investigated $Q/2\pi=(\sqrt{5}+1)/2$ and checked the qualitative features for several other incommensurate wavevectors.
The general features are:
\begin{enumerate}
	\item The extended states on the clean ($A_J=0$) axis persist in the presence of small modulation $A_J$. With increasing $A_J$, the high energy states localize before the lower energy states.

	\item Near the Ising transition along the phase boundary above the diagonal $A_J = J$, the gap closes and reopens linearly and all excitations are extended up to a finite energy \emph{above} the gap. See the first two columns of Fig.~\ref{fig:cuts-enipr} for representative spectra and IPR behavior and Sec.~\ref{sec:GroundStateTransitions} for more discussion.

	\item For $A_J/h>2$, all states are localized, consistent with the analytically proven behavior at $J=0$ (Sec.~\ref{sub:pure_modulation_aaa_triality}) and the analysis at large $A_J$ (Sec.~\ref{sub:large_amplitude}) and large $J$ (Sec.~\ref{sub:large_Jlimit}).  
	At large $J/h$, we observe the expected energy independent localization transition near $A_J/h = 1$. 

	\item On the $J=0$ line at $A_J/h < 2$, we confirm that all states are critical by calculating the scaling of the IPR. At high energy, the states localize on the introduction of $J>0$. However, we find that the lowest energy states (above the gap) continue to exhibit critical IPR scaling throughout the triangle below the diagonal $A_J = J$, see representative data in Fig.~\ref{fig:cuts-enipr}. This behavior defines the purple region in Fig.~\ref{fig:phasediagram}.

\end{enumerate}



\section{Ground state quantum phase transitions}
\label{sec:GroundStateTransitions}

In this section, we focus on the properties of the zero temperature quantum phase transition between the ground state para- and ferro-magnetic phases.
The phase transitions above and below the diagonal $A_J = J$ in Fig.~\ref{fig:phasediagram} are qualitatively distinct.
Above the diagonal ($A_J < J$), the quasiperiodic modulation is irrelevant and the transition lies in the standard 1D quantum Ising universality class. 
Below the diagonal ($A_J > J$), the transition lies in a new `quasiperiodic Ising' universality class with behavior intermediate between the clean Ising critical point and the infinite randomness critical point (IRCP) which governs the disordered Ising transition \cite{Fisher:1995oq}. 

At both transitions, the correlation length $\xi$ diverges with exponent $\nu = 1$. 
From the relation $\xi \sim -1/I$, where $I$ is the integral governing the convergence of the Majorana boundary mode, Eq.~\eqref{Eq:IIntegral}, it is straightforward to show that: 
\begin{align}
	 \label{Eq:XiDivergenceGS}
	 \xi \sim \delta^{-1},
\end{align}
where $\delta$ is the deviation from the phase boundary at $I=0$. 
This is consistent with the Harris-Luck criterion \cite{Luck:1993fu,Luck:1993ad}, which imposes that $\nu \ge 1/d = 1$ for phase transitions in the presence of incommensurate modulation.

\subsection{Clean Ising Transition} 
\label{sub:clean_gs_transition}

While $\nu = 1$ at both transitions, the dynamical properties are quite distinct. 
Above the diagonal, the phase boundary connects to the standard clean Ising transition at $J/h = 1, A_J/h=0$. 
At small $A_J$, as the Harris-Luck criterion is marginal, it is natural to conjecture that weak quasiperiodic modulation is (marginally) irrelevant. 
This would imply that in the vicinity of the phase boundary
\begin{enumerate}
	\item the dynamical critical exponent $z=1$ so that the gap $\Delta$ closes linearly with $\delta$: 
\begin{align}
	\Delta \sim \delta^{\nu z} \sim \delta;
\end{align}

	\item the low-energy excitations are extended, as these are the excitations which mediate the phase transition.
\end{enumerate}
Both of these expectations are borne out numerically quite beautifully.

\begin{figure*}[t]
	\centering
	\includegraphics[width=\textwidth]{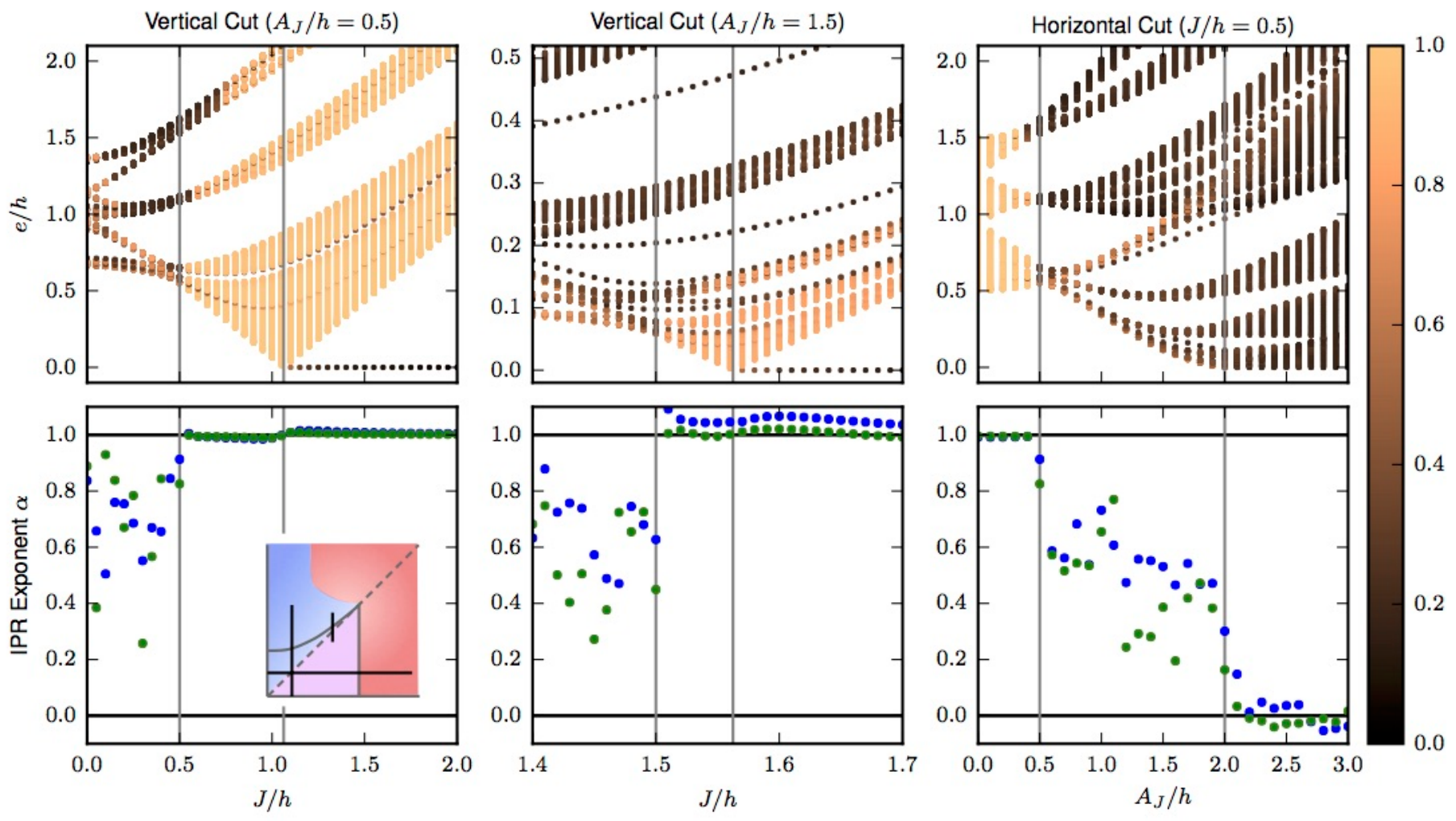}
	\caption{%
	Fermionic excitation spectra (upper row) and scaling exponent of low energy IPR (lower row) on representative cuts in the phase diagram, indicated by black lines on the inset phase diagram. 
	On each panel, vertical lines indicate the ground state symmetry breaking phase boundary and the diagonal ($A_J = J$) critical to extended boundary.
	(upper row) 
	Excitation spectra at size $L=1000$. 
	The color value of each level indicates the extension of the corresponding state. 
	The color value is given by $- \log \textrm{IPR} / \log L$, which varies between 0 (localized, black) and 1 (extended, copper).
	(lower row)
	The IPR exponent $\alpha$ is defined by the scaling $[\log IPR] \sim -\alpha \log L$, where the mean $[ \cdot ]$ is taken over the lowest twentieth of the excitations.
	$\alpha$ varies between $0$ (localized) and $1$ (extended).
	To indicate the finite-size trends approaching the thermodynamic limit, we fit data from both $L=100,200,350,500,750,1000$ (blue) and from $L=350,500,750,1000$ (green).
	}
	\label{fig:cuts-enipr}
\end{figure*}

The top left panel of Fig.~\ref{fig:cuts-enipr} shows the excitation spectrum versus $J/h$ for a vertical cut at $A_J/h=0.5$.
The vertical cut intersects three regimes: the critically delocalized  PM ($0 \le J/h \le 0.5$), the extended PM ($0.5 \le J/h \le J_c/h = 1.0625$), and the extended FM ($J_c/h \le J/h$).
The boundary zero mode associated with the FM is clearly visible to the right of the transition.
As promised, the gap closes linearly and the low-energy excitations above the gap remain extended.
The extension of the wavefunctions are indicated qualitatively by the coloring of the states; quantitatively, the lower panel shows the scaling exponent $\gamma$ of the low energy IPR with system size,
\begin{align}
	\mathrm{IPR} \sim 1/L^{\gamma}
\end{align}
The lower left panel shows that $\gamma$ is near $1$ for the low energy excitations for $J/h>0.5$ and in particular near the transition.

The center column of Fig.~\ref{fig:cuts-enipr} presents similar data at stronger incommensurate modulation, $A_J/h = 1.5$.
Note that the axes are zoomed in on the phase boundary and on low energies relative to the previous plot.
The high-energy excitations ($e/h>\approx 0.2$) are localized across the transition.
Nevertheless, the gap closes linearly (center top) and the low-energy excitations remain extended nearby ($\gamma \approx 1$, in the bottom panel).
Thus, the ground state symmetry-breaking transition is still in the clean quantum Ising universality class. 

\subsection{Quasiperiodic Ising Transition} 
\label{sub:qp_gs_transition}

Below the diagonal, the dynamics near the transition between the QPFM and PM change character rather dramatically. 
Along the $J = 0$ line, the symmetry breaking transition at $A_J/h = 2$ coincides with the transition from critically delocalized to localized excitations at all energies, as shown in Sec.~\ref{sub:pure_modulation_aaa_triality}. 
There are no fully extended excitations.
In analogy with the irrelevance of $A_J$ at the clean transition, we conjecture that $J$ is irrelevant at low energies to the strong quasiperiodically driven transition along the vertical phase boundary.

Numerically, this conjecture is borne out by the following observations (see third column of Fig.~\ref{fig:cuts-enipr} for representative data  along a particular cut at $J/h = 0.5$):
\begin{itemize}
	\item The single particle gap closes on approaching the transition from the paramagnetic side with an exponent $z \approx 2$. However, the gap does not reopen on the symmetry breaking side of the transition.

	\item All excitations are localized in the symmetry breaking phase at $A_J/h > 2$.

	\item The low energy excitations (bottom tenth of states above the gap) in the paramagnetic phase exhibit critical IPR scaling and extreme finite size sensitivity. See the lower row of Fig.~\ref{fig:cuts-enipr}. The IPR exponent $\gamma$ lies properly between $0$ and $1$ and exhibits strong finite-size fluctuations.

	\item At higher energy (and $J/h > 0$), the excitations can be localized even in the paramagnetic phase. We do not know whether the critical low energy states are separated from these localized states by a mobility edge or whether there is a long crossover.

\end{itemize}

While we leave a full analytic study of this transition to forthcoming work \cite{UsSoon}, it is clear that the quasiperiodic Ising transition is intermediate between the well-known clean and infinite randomness critical points. 
In the former, the critical excitations are extended on both sides of the transition with dynamical exponent $z=1$; while in the latter, the excitations are localized on both sides of the transition and the dynamics are activated (roughly, $z\to\infty$). 
Across the quasiperiodic transition, the excitations pass from critically delocalized and gapped (with exponent $z=2$) to localized and gapless across the transition.
Also, unlike the infinite randomness transition, the correlation functions in the quasiperiodic case do not acquire broad distributions at long distances.
For example, the mean and typical decay of the boundary mode are governed by the same correlation length $\xi$ with exponent $\nu = 1$ while in the infinite randomness transition these are governed by two distinct diverging length scales (with $\nu = 2$ and $1$, respectively).

\section{Excited state Ising glass order}
\label{sec:excited_glass_order}

We expect localized Ising glass order at all energy densities in the regions where 
\begin{enumerate}
	\item the ground state breaks Ising symmetry, and
	\item all the excitations are localized. 
\end{enumerate}
This region is indicated in red in Fig.~\ref{fig:phasediagram}. 
Note that the boundary of the region above the diagonal ($J>A_J$) depends on $Q$.
Although we do not have an explicit functional form for this $Q$ dependence, the analysis in Sec.~\ref{sec:localization_of_excitations} constrains the boundary to be at $A_J/h=1$ as $J/h \to \infty$ independent of $Q$.

\begin{figure*}[t]
	\centering
	\includegraphics[width=\textwidth]{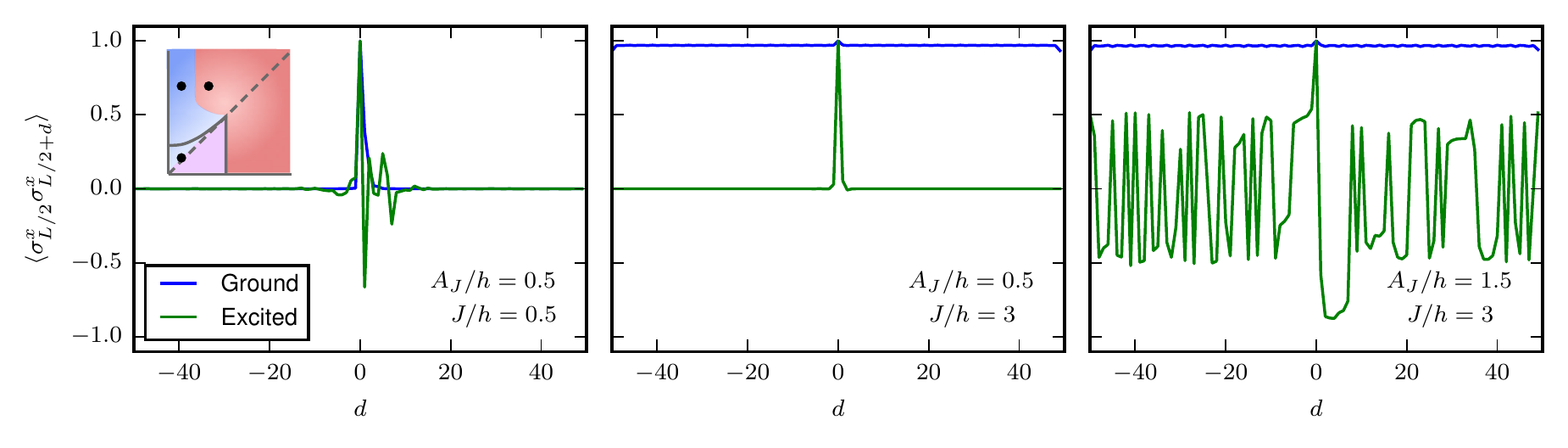}
	\caption{Spin-spin correlation function $\langle \sigma^{x}_{L/2}\sigma^x_{L/2+d}\rangle $ as a function of the distance $d$ from the center of the chain in a $L=100$ site chain with open boundary conditions. 
	Blue: ground state, green: a random excited state from the infinite temperature ensemble. 
	From left to right, the parameters are chosen so that the system is critical paramagnetic, extended ferromagnetic and localized ferromagnetic, as indicated in the inset.}
	\label{fig:correlationfuncs}
\end{figure*}

\emph{Representative correlation functions:}
Fig.~\ref{fig:correlationfuncs} shows the spin-spin correlation function $\langle\sigma^{x}_{L/2}\sigma^x_{L/2+d}\rangle$ in the ground state (blue) and a random excited state drawn from the infinite temperature ensemble (green) at three different representative points in the phase diagram.
In the left panel, the spin-spin correlation function decays rapidly in both states, confirming that the point $A_J/h=J/h=0.5$ is in the PM phase. 
The center panel shows the correlation functions at a point in the phase diagram where the ground state is ordered and the excitations above it are extended. The long-range order in the ground state is clearly detected by the blue curve which approaches a non-zero value at long distance $d$.
The green curve, on the other hand, decays quickly to zero and shows that infinite temperature eigenstates are not ordered. 
This agrees with the Mermin-Wagner-Peierls theorem that states there can be no long-range order at any finite temperature in $1D$.

The right panel provides evidence for Ising glass order in a randomly chosen infinite temperature state at a point in the phase diagram where the ground state is ordered and the single-particle excitations are localized. 
The green curve does not decay as $d \to \infty$; instead it fluctuates on an order one scale depending on whether spin $L/2$ and spin $L/2+d$ are aligned or anti-aligned in the chosen infinite temperature state. 

\emph{Frozen order parameter:}
The quasiperiodic Ising glass order is detected by an order parameter which generalizes the ground state order parameter, Eq.~\eqref{eq:gsorderparameter}, to finite energy density states:
\begin{align}
	\label{eq:excitedorderparameter}
 	\mathcal{O}_{exc} = [|\langle E| \sigma^x_{L/4+j} \sigma^x_{3L/4+j}|E\rangle|]_{E,j}
\end{align} 
Here, $[ \cdot ]_{E,j}$ indicates averaging over both eigenstates $\ket{E}$ in some energy window and over a small spatial window $j \in [-w, w]$ to suppress the quasiperiodic fluctuations.
$\mathcal{O}_{exc}$ is non-zero as $L\to \infty$ only in states with long range Ising symmetry breaking, such as the Ising glass.
We plot the order parameter along several representative cuts in the phase diagram in Fig.~\ref{fig:frozenmag} at $L=500$ for the ground state (blue) and states drawn from the infinite temperature ensemble (green).

The three panels are consistent with the Ising glass order being present only in the red shaded region. 
In the rightmost panel, the weak quasiperiodic modulation leaves the excitations extended at all $J/h$; accordingly, the excited states are always paramagnetic, irrespective of ground state ordering.
In the center panel at large $J/h$, the excited state order develops as the excitations localize across the AA-like transition described in Sec.~\ref{sub:large_Jlimit}.
The ground state order parameter is completely insensitive to the excited state ordering.
Finally, in the leftmost panel, the ground and excited state order develop at the same coupling $A_J/h=2$ because the ground state symmetry breaking phase transition and the localization transition of the excitations coincide, as discussed in Sec.~\ref{sub:qp_gs_transition}.

\subsection{Excited state transition} 
\label{sub:excited_state_transition}

The Ising glass order is destroyed if either the ground state becomes paramagnetic or the domain wall excitations delocalize.
The central panel of Fig.~\ref{fig:frozenmag} illustrates the latter transition at large $J/h$. 
In the leftmost panel of the same figure, on the other hand, the ground state transition coincides with the delocalization of excitations, see Sec.~\ref{sub:qp_gs_transition}.
Below, we develop a picture of the transition of the central panel, which is entirely governed by the localization properties of the excitations. 
We leave more careful study of the other excited state transition to future work \cite{UsSoon}.

At large $J/h$, the ground state magnetization $M^0_i$ is very close to one on each site $i$. 
In the excited states, the magnitude of the magnetization is reduced by the fluctuations of domain walls across site $i$.
When the domain walls have typical localization length $\xi$, there are $\sim \xi$ such relevant domain walls.
Any domain wall localized further away merely flips the sign of $M_i$ without reducing its magnitude.
Thus, the fluctuations of the parity of the domain walls to the left of $i$ ultimately control the excited state magnetization.

Mathematically,
\begin{align}
	|M_i| &= |\langle \sigma^x_i \rangle| \approx |M^0_i| \,|\langle (-1)^{N_{<i}}\rangle| 
\end{align}
where $N_{<i}$ is the number of domain walls to the left of site $i$. 
As the domain walls are non-interacting, we have
\begin{align}
	N_{<i} &= \sum_{\alpha} n_\alpha \mathbb{I}[\textrm{DW } \alpha \textrm{ at position} < i]
\end{align}
where $n_\alpha$ is the occupation of domain wall eigenstate $\alpha$ and the indicator function is 1 if that domain wall is to the left of $i$.
Within an excited eigenstate, $N_{<i}$ is thus a sum of independent random variables with mean
\begin{align}
	\langle N_{<i} \rangle = \sum_\alpha n_\alpha P^\alpha_{<i}
\end{align}
and variance
\begin{align}
	\langle (\delta N_{<i})^2\rangle &= \sum_\alpha n_\alpha P^\alpha_{<i}(1-P^\alpha_{<i})
\end{align}
Here, $P^\alpha_{<i}$ is the probability that the domain wall in state $\alpha$ lies to the left of $i$.
As the mean value of $N_{<i}$ only adjusts the overall sign of $M_i$, we focus on the fluctuations $\delta N_{<i}$ to estimate the reduction of $|M_i|$.
In the localized regime, only those eigenmodes $\alpha$ with localization centers within a distance $\xi$ of $i$ contribute to these fluctuations as $P^\alpha_{<i}$ approaches $0$ or $1$ further away.

At large $\xi$, $\delta N_{<i}$ becomes a Gaussian distributed random variable with variance $\propto \xi$. 
The magnitude of the magnetization is thus reduced by,
\begin{align}
	|\langle (-1)^{N_{<i}}\rangle| &\approx |\langle e^{i \pi \delta N_{<i}}\rangle|\sim e^{-a \xi}
\end{align}
where $a$ is related to the proportionality constant in the variance. 
Since in the large $J/h$ limit, the localization transition is of AA-type, we have $\xi \sim \delta^{-1}$ (see Sec.~\ref{sub:large_Jlimit}). 
This leads finally to an essentially singularity in the excited state order at the transition,
\begin{align}
	|M| \sim e^{-a'/\delta}
\end{align}
with $a'$ a $\delta$-independent constant.

The quasiperiodic Ising order parameter $\mathcal{O}_{exc}$ of Eq.~\eqref{eq:excitedorderparameter} should approach $|M|^2$ at large $L$ (and large spatial averaging window $w$) and accordingly inherits the essential singularity.
Our numerics are consistent with this form but are inconclusive as it is difficult to distinguish an essential singularity from a shift in the apparent critical point.

\section{Conclusions}
\label{sec:Conclusions}

We have presented the first analytical study of localization-protected excited state order without disorder. 
Incommensurate modulation of the exchange couplings leads to a large Ising glass phase in the canonical quantum Ising chain.
By arguments similar to those presented in the context of disordered Ising chains \cite{Huse:2013aa}, we expect that the glass survives the introduction of weak interactions, so long as the localization length $\xi$ of the domain walls is sufficiently short compared to the typical domain wall density. 
Quasiperiodic modulation arises naturally in optical experiments; it also provides an analytic platform for further study of localization.
In some ways, it is simpler than disordered localization, which is plagued by rare-region effects and Griffiths' phases \cite{Vosk:2013yg,Potter:2015ab,Agarwal:2015aa,Gopalakrishnan:2016aa,Khemani:2016aa}. 
Understanding the nature of the quasiperiodic localization transition, with and without interactions, may thereby cut to the heart of the phenomenon.

We also presented a theory of the melting transition for the excited state order in the non-interacting case. 
Reducing the amplitude of modulation leads to delocalization of the domain walls which may be accompanied by a ground-state symmetry-breaking transition.
The divergence of the localization length $\xi$ of domain walls leads to an essential singularity in the excited state order. 
However, as interactions are likely to delocalize the system before $\xi$ diverges, we expect the essential singularity to be cutoff and the transition to qualitatively change.
This is an especially interesting direction for future work.

Our analysis is aided by our generalization of Aubry-Andr\'e duality, `AAA triality', which applies along the pure modulation axis ($J = 0$) of the phase diagram in Fig.~\ref{fig:phasediagram}. 
The Ising-symmetry breaking ground state transition at $A_J/h=2$ coincides with the `self-trial' point of this transformation.
The triality requires that the wavefunctions are critically delocalized at all single-particle energies on one side of the transition (PM) and localized on the other (QPFM).
The associated quantum critical point thus lies neither in the clean Ising universality class where the excitations are extended, nor in the infinite randomness class of disordered systems where the excitations are fully localized.
Intriguingly, numerics show that these critical properties persist away from the pure modulation line at low energies.
This suggests that the entire phase boundary lies in this intermediate universality class, the `quasiperiodic Ising class' \cite{UsSoon}.

From a zero-temperature perspective, we expect the three ground state phases to be stable to the inclusion of interactions as they are either gapped or localized. 
Whether interactions are irrelevant at the peculiar critical point discussed in the previous paragraph is an interesting open question.
The transition is neither clean enough for a field theoretic renormalization treatment \cite{Cardy:1996aa} nor disordered enough to obviously flow to infinite randomness under real space renormalization \cite{Fisher:1995oq}.
Perhaps a coarse-graining treatment could be developed in the semi-classical limit of $Q \to 0$, as has been done for the AA model \cite{Wilkinson:1984aa}.

Finally, we would like to comment on the potential for studying the quasiperiodic Ising glass experimentally. 
Essentially, any quantum optical system that realizes a tunable Ising chain would be able to probe the excited state Ising glass order by quench experiments. 
These include, for example, linear chains of trapped ions using hyperfine states as Ising degrees of freedom \cite{Islam:2011aa,Smith:2016aa}, Rydbergs trapped in optical tweezers \cite{Glaetzle:2015aa,Labuhn:2016aa}, chains of trapped ions undergoing the zig-zag transition \cite{Enzer:2000aa,Shimshoni:2011aa} or ultracold atoms  undergoing a staggering transition in a tilted lattice potential \cite{Simon:2011fk}.
The simplest way to apply an incommensurate modulation in these systems is to modulate the position of the atoms/ions using an extra effective spatial potential, whether that be with a standing wave or with optical tweezers.

However, there are two classes of Ising model simulators: those where the Ising degree of freedom is spatial (such as at the zig-zag transition) and those where it is internal (such as in the ion trap Ising simulator). 
In the former, as the order parameter directly couples to spatial position, incommensurate modulation potentials locally break the Ising symmetry. 
This raises interesting questions regarding the Imry-Ma stability of excited state symmetry-breaking order \cite{Imry:1975aa}, which we leave for future work.
In the latter Ising simulators, modulating the position can directly modulate couplings without introducing Ising odd terms (ie.  effective longitudinal fields). 
We expect these platforms to be able to realize the quasiperiodic Ising glass directly.


\begin{acknowledgments}
We would like to thank V. Oganesyan, D. Huse, A. Polkovnikov and A. Kaufman for helpful discussions. 
We thank the Kavli Institute for Theoretical Physics (KITP) in Santa Barbara for their hospitality during the early stages of this work and the National Science Foundation (NSF) under Grant No. NSF PHY11-25915 for supporting KITP.
C.R.L. acknowledges support from the Sloan Foundation through a Sloan Research Fellowship and the NSF through Grant No. PHY-1656234. 
Note that any opinion, findings, and conclusions or recommendations expressed in this material are those of the authors and do not necessarily reflect the views of the NSF.
\end{acknowledgments}

\bibliography{paper-master}

\appendix

\section{Ground state phase boundary} 
\label{app:ground_state_phase_boundary}

We include the elementary derivation of the contours of $I$ at $A_h=0$.
Consider the differential (setting $h= 1$ and assuming $A_J, J > 0$ in order to avoid writing absolute values throughout):
\begin{align}
	dI &= - \int_0^{2\pi} \frac{d\theta}{2\pi} \frac{d J + dA_J \cos(\theta)}{ J + A_J \cos(\theta)}
\end{align}
Using contour integration on the unit circle $z = e^{i \theta}$,
\begin{align}
	dI &= -\frac{2}{A_J} \oint \frac{dz}{2\pi i} \frac{d J + dA_J (z + z^{-1})/2}{(z-z_+)(z-z_-)}
\end{align}
where $z_\pm = -\frac{ J}{A_J} \pm \sqrt{\left(\frac{ J}{A_J}\right)^2 - 1}$. 
For $ J > A_J$, the pole at $z_+$ lies inside the unit circle while for $ J < A_J$, both $z_+$ and $z_-$ lie outside. Explicitly,
\begin{align}
	\begin{pmatrix}\frac{\partial I}{\partial  J} \\
    \frac{\partial I}{\partial A_J}
    \end{pmatrix} &= \left\{ \begin{matrix} 
    	\begin{pmatrix} 
          \frac{1}{\sqrt{ J^2 - A_J^2}} \\
          1/A_J - \frac{2  J/A_J}{\sqrt{ J^2 - A_J^2}}
        \end{pmatrix} &  J > A_J \\
    	\begin{pmatrix} 
          0 \\
          1/A_J 
        \end{pmatrix} &  J < A_J 
    \end{matrix}\right.     
\end{align}
Thus, the contours of $I$ behave differently in the regions above and below the diagonal $ J = A_J$. 
Solving $dI = 0$ above the diagonal leads to parabolic contours of the form,
\begin{align}
	 J = J_0 + \frac{A_J^2}{4 J_0}
\end{align}
Below, the contours are vertical. 
The functional form of $I$ then follows immediately from explicit integration of \eqref{Eq:IIntegral} on the axes:
\begin{align}
	I &= \left\{ \begin{matrix}
    -\log |J_0| = - \log \left| \frac{ J+ \sqrt{ J^2 - A_J^2}}{2} \right| 	&  J > A_J \\
	-\log |A_J|/2 		&  J < A_J
	\end{matrix} \right.
\end{align}


\section{General commensurate analysis} 
\label{app:general_commensurate_analysis}

In the appendices, we provide a detailed analysis of the spectral properties of the excitations in the TFIM with \emph{commensurate} modulation $Q = 2\pi p/q$, for $p$ and $q$ coprime integers. 
The incommensurate TFIM then arises by taking the limit $p,q \to \infty$ such that $Q/2\pi$ approaches an irrational value. 
For example, to study the incommensurate TFIM modulated with a wavevector corresponding to the golden mean, $Q/2\pi = (1+\sqrt{5})/2$, one could take the sequence $p_n/q_n = F_n/F_{n-1}$, where $F_n$ is the $n$'th Fibonacci number.
Thouless \cite{Thouless:1983aa,Han:1994cs} emphasized this approach to studying the original Aubry-Andr\'e model, pointing out the $q$ plays a role analogous to finite size in a scaling theory of the transition, as it determines the length over which the incommensurate model can be approximated by the commensurate one. 
Here, we review and generalize the commensurate analysis of Ref.~\cite{Han:1994cs} to the TFIM.

For commensurate modulation $Q/2\pi=p/q$, Bloch's theorem implies that the single-particle energy spectrum of Eq.~\eqref{Eq:MajoranaHam} is a function of the quasi-momentum $k_x \in [-\pi/q, \pi/q)$ with $2q$ bands. 
The 1D bands depend on the phase $\phi$ through the explicit Hamiltonian dependence on $\phi$; we note that this dependence is periodic in $\phi \to \phi + 2\pi / q$, as the site labels can be shifted by an integer $j \to j - l$ where $lp = 1 (\textrm{mod } q)$ to absorb such a shift. 
(As $p$ and $q$ are coprime, $p$ has a multiplicative inverse modulo $q$.)
It is natural to view the bands as 2D sheets over both $k_x$ and $\phi$ (this is the band structure of the associated 2D model, Eq.~\eqref{Eq:2DHofstadterHam}), but one must remember that the bands of the actual 1D model corresponds to a $k_x$ slice of the 2D bands at fixed $\phi$. 

Explicitly, the eigenvalue problem of Eq.~\eqref{Eq:1DComplexFermionModel} satisfies the difference equations:
\begin{align}
\label{Eq:DiffEqns}
-iJ_{i-1/2} \psi_{2i-1} + i h_i \psi_{2i+1} &= E \psi_{2i} \nonumber\\
-i h_i \psi_{2i} + iJ_{i+1/2} \psi_{2i+2} &= E \psi_{2i+1}
\end{align}
where $h_i, J_{i+1/2}$, defined in Eq.~\eqref{Eq:HcouplingDef}, are periodic with period $q$.
By Bloch's theorem, the solutions satisfy the twisted boundary conditions on a $2q$-site chain:
\begin{align}
	\psi_{2(n+q)} &= e^{i k_x q} \psi_{2n} \nonumber\\
    \psi_{2(n+q) + 1} &= e^{i k_x q} \psi_{2n + 1}
\end{align}
The eigenspectrum of the infinite set of equations in Eq.~\eqref{Eq:DiffEqns} follow from the eigenspectrum of the finite matrix
\begin{align*}
&M(k_x,\phi) = \nonumber \\
&\left(\begin{array}{cccccc}
0 &ih_0 &0 &0&\ldots &-i J_{-\frac{1}{2}} e^{-ik_x q} \\
-ih_0 & 0 &iJ_{\frac{1}{2}}& 0 & \ldots & 0 \\
0 & -iJ_{\frac{1}{2}} & 0 &ih_1& \ldots & 0 \\
\vdots & & \vdots &  & \vdots & \\
i J_{-\frac{1}{2}}e^{ik_x q} & 0 & 0 &0& \ldots& 0
\end{array}\right)
\end{align*}
with $J_{q-1/2}= J_{-1/2}$. 
The characteristic polynomial of $M$ is:
\begin{align}
C(E; k_x, \phi) = \textrm{det}[M - E I]
\end{align}
In general, the characteristic polynomial depends on the phase $\Delta$ as well; we suppress this dependence as we specialize to the case $A_h=0$ below.
The analysis can be straightforwardly extended to non-zero $A_h$.

$C(E;k_x,\phi_h)$ is a polynomial of $E$ of degree $2q$.
Since $M$ is Hermitian, the polynomial is real (for $E$ real):
\begin{align}
C(E; k_x, \phi) = C^*(E; k_x, \phi)
\end{align}
The bipartite (chiral) symmetry $\mathcal{C}$ which takes $\psi_{j} \to (-1)^j \psi_j$ anticommutes with $M$: $\mathcal{C} M \mathcal{C} = - M$. 
Thus, the characteristic polynomial is an even function of $E$,
\begin{align}
C(-E; k_x, \phi) = C(E; k_x, \phi),
\end{align}
and accordingly that the eigenvalues come in $\pm E$ pairs within each $k_x$ sector.

By explicit evaluation, $C$ has the following structure,
\begin{align}
\label{Eq:Cek1k2}
C(E; k_x, \phi) = \sum_{m=0}^{q} K_m(\phi) E^{2m} - 2(-1)^{q}P(\phi)\cos(k_x q)
\end{align}
where all of the $k_x$ dependence lies in the the energy independent term. 
This dependence follows from the two $k_x$ dependent terms in the explicit expansion of the determinant:
\begin{align}
(-1)^{2q-1} e^{ik_xq} \prod_{j=0}^{q-1} (ih_j) \prod_{m=0}^{q-1} (iJ_{m+1/2}) + c.c.
\end{align}
which simplifies to the last term in Eq.~\eqref{Eq:Cek1k2} using $h_j=1$ and the definition,
\begin{align}
P(\phi) \equiv \prod_{j=0}^{q-1} ( J+A_J \cos(2\pi p (j+1/2)/q + \phi))
\end{align}
In Appendix~\ref{app:EvalPk2}, we evaluate this to be
\begin{align}
P(\phi) = 2 \left(\frac{A_J}{2}\right)^q \left( T_q( J/A_J) - (-1)^{q+p} \cos(q\phi) \right)
\label{Eq:Pk2FinalExpMainText}
\end{align}
where $T_q$ is the Chebyshev polynomial of order $q$.

Next, we evaluate the rest of the constant term, $K_0(\phi)$. This is most easily accomplished by working at $k_x = E = 0$, where $M$ is antisymmetric and the determinant is the square of the Pfaffian. At $h_j=1$, we have:
\begin{align}
K_0(\phi) 
&= (-1)^q [(P(\phi))^2 + 1] \label{Eq:K0k2}
\end{align}

To make further progress, we need control of the higher order coefficients $K_m(\phi)$. 
For general couplings, these are not so easy to compute, although we setup some formalism exploring this in App.~\ref{app:higher_order_coefficients}.
In the important special case of $J=0$, the triality discussed in Sec.~\ref{sub:pure_modulation_aaa_triality} allows us to show that the $K_m$ are actually independent of $\phi$ for all $m > 0$, see App.~\ref{app:higher_order_coefficients_with_triality}. 
With this simplification, we will be able to determine the energy independent localization properties on the $J=0$ line, see App.~\ref{app:localization_at_j0}.

\section{Higher order coefficients with triality} 
\label{app:higher_order_coefficients_with_triality}

In the Aubry-Andre models \cite{Aubry:1980aa, Thouless:1983aa, Han:1994cs}, the AA duality implies that $C(E;k_x, \phi;t,V) = C(E;\phi, -k_x; V,t)$. Equating this order by order in $E$, 
\begin{align}
	K_m(\phi;t,V) &= K_m(-k_x; V,t)
\end{align}
for all $m>0$. 
Differentiating with respect to $\phi$, we see that $K_m(\phi)$ is independent of $\phi$ for $m>0$.
This leads to the energy independence of the AA localization transition by the logic described in App.~\ref{app:localization_at_j0}.

In the AAA models defined by Eq.~\ref{eq:HamiltonianAAA}, the triality transformation implies that $C(E; k_x, \phi; A_1, A_2, A_3) = C(E; R(k_x, \phi); A_2, A_3, A_1)$, where $R$ is the linear transformation implementing the three-fold rotation on the momentum space defined by $k_x, \phi$. 
We note that $R$ is a geometric rotation by $2\pi/3$ conjugated by an anisotropic scale transformation as our embedding of the honeycomb structure is into a rectangular lattice as in Fig.~\ref{fig:TFIMcouplings}c, rather than the  geometrically symmetric embedding in Fig.~\ref{fig:honeycomblayers}.
Nonetheless, 
\begin{align}
	K_m(\phi; A_1, A_2, A_3) &= K_m( \alpha k_x + \beta \phi; A_2, A_3, A_1)
\end{align}
where $\alpha, \beta$ are the appropriate matrix elements of $R$. 
Again, differentiating with respect to $k_x$, it follows that $K_m$ is actually independent of it's first argument $\phi$ in the AAA models. 

In particular, this holds along the $J=0$ line of the incommensurate TFIM, which corresponds to $A_3 = h, A_1 = A_2 = A_J/2$.

\section{General higher order coefficients} 
\label{app:higher_order_coefficients}

In general in the TFIM, the coefficients $K_m(\phi)$ are both $\phi$-dependent and non-trivial to evaluate for $m > 0$. Although we do not need any of the following formalism for the results used in the manuscript, we summarize here a few formulae for posterity.

To calculate the higher order terms it is helpful to consider the explicit representation of the determinant (with $B = M - E$),
\begin{align}
	\det B = \sum_{\pi \in S_{2q}} (-1)^\pi B_{1,\pi_1} B_{2,\pi_2} \cdots B_{2q,\pi_{2q}}.
\end{align}
The factors of $E$ come from diagonal matrix elements, so the coefficient $K_m$ comes from the permutations which hold $2m$ sites fixed but are otherwise off-diagonal. 
\begin{align}
	K_m &= \sum_{i_1<i_2< \cdots < i_{2m}} \sum_{\myover{\pi \in S_{2q}}{\pi_{i_l} = i_l}} (-1)^{\pi} \prod_{i \notin i_l} M_{i,\pi_i}
\end{align}
Moreover, since $M$ only connects adjacent sites, only those permutations which permute within each diagonal block (from $i_j + 1$ to $i_{j+1} - 1$) are non-zero. 
At fixed $i_1 \ldots i_{2m}$, the contribution to $K_m$ thus factors into a product of the determinants of the diagonal subblocks of $M$ between rows $i_l$ and $i_{l+1}$: 
\begin{align}
	K_m &= \sum_{i_1<\cdots<i_{2m}} \prod_{j=1}^{2m} \det M\vert_{i_j+1, i_{j+1} - 1}
\end{align}
where the $l=2m$ block wraps around the corners of the matrix. 
Since each subblock of $M$ is antisymmetric, the determinant is non-zero only if $i_{j+1} - i_j - 1$ is even -- that is, if the $i_j$ alternate between even and odd. 
Explicitly, 
\begin{align}
	\det M\vert_{i+1,j-1} &= i^{j-i-1}\left\{ \begin{array}{ll}  \prod_{l = (i+1)/2}^{j/2-1} h_l^2 & i/j\textrm{ odd/even} \\
    \prod_{l=(i+1)/2}^{(j/2-1} J_l^2 & i/j \textrm{ even/odd} \end{array} \right.
\end{align}
Keep tracking of the number of $i$'s,  we find that the coefficients $K_m$ alternate in sign.

For $A_h = 0$,
\begin{align}
	 \det M\vert_{i+1,j-1} = i^{j-i-1} h^{j- i - 1} e^{-V(i,j)}
\end{align}
where 
\begin{align}
\label{Eq:VijDomainWalls}
	V(i,j) = \left\{ \begin{array}{ll}
    - \sum_{l = (i+1)/2}^{j/2-1} 2\log (J_l/h)  & i/j\textrm{ even/odd} \\
    0	& i/j \textrm{ odd/even} \\
    \infty & \textrm{else}
    \end{array} \right.
\end{align}
In this form, $K_m$ looks like,
\begin{align}
	K_m &= (-1)^{q-m} h^{2q - 2m} \sum_{i_1 < \cdots < i_{2m}} e^{-\sum_{j=1}^{2m} V(i_j, i_{j+1})} \label{Eq:KmPartitionSum}
\end{align}
The magnitude of $K_m$ is given by the canonical partition sum at temperature one for a system of length $2q$ with $2m$ domains of energy $V(i,j)$. 

The number of wavelengths of $J$ which fit in the chain of length $2q$ is $p$. If $p$ is held fixed as $q \to \infty$, then $J_{i+1/2}$ is smooth on the lattice scale $a$, as the wavelength of the incommensurate field $q/p \gg a$. In this limit, the continuum approximation in which $a=0$ ought to be a good starting point for a semi-classical analysis that incorporates the leading effects of a small $a/q$. If on the other hand, $p/q$ approaches a non-zero constant as $q \to \infty$ (for example, the golden mean), then $q/p \sim O(a)$. 
We leave further analysis of $K_m(\phi)$ in these limits to future work.

\section{Localization properties at $J=0$} 
\label{app:localization_at_j0}

From App.~\ref{app:higher_order_coefficients_with_triality}, the characteristic polynomial may be written along the $J=0$ axis,
\begin{align}
	\label{Eq:Cnophi}
C(E; k_x, \phi) &= \sum_{m=1}^{q} K_m E^{2m} \nonumber \\
&+ (-1)^q( P(\phi)^2 - 2P(\phi)\cos(k_x q)+1)
\end{align}
where $K_m$ depends on couplings $A_J$, $h$ but not on $k_x$ or $\phi$. 

This form entails a remarkable geometric property of the $2q$ bands in the 2D bandstructure: they are all approximately the same shape up to shift and scale. 
This implies that the \emph{ratio} $r$ of the $k_x$ and $\phi$ dispersion of each band is independent of band number $n$. Thus, if this ratio approach zero sufficiently rapidly with $q$ as we approach the incommensurate limit by $Q = 2\pi p /q$, the total dispersion in the $k_x$  must also go to zero even summed over all $q$ bands (because the total 2D spectrum is bounded). 
This argument allows us to show that the spectrum is fully localized for $A_J > 2$. 
For $A_J < 2$, the $r$ approaches a constant as $q \to \infty$, which indicates the presence of critical states, also independent of energy.
See \cite{Han:1994cs} for similar analysis in a simpler model.

Suppose we identify the $n$'th zero of $C(E)$ at fixed $(k_x, \phi) = (0,0)$:
\begin{align}
C(E_n^0; 0,0)=0
\end{align}
The dispersion of the $n$'th band is then determined by following this zero while varying $k_x$ and $\phi$. To linear order in $E - E_n^0$, we have
\begin{align}
	0 &= C(E_n^0; k_x, \phi) + (E_n(k_x,\phi) - E_n^0) C'_n
\end{align}
where $C'_n = \left. \frac{\partial C}{\partial E}\right|_{E_n^0}$ is a constant independent of the 2D momenta from Eq.~\eqref{Eq:Cnophi}. 
Rearranging, this implies the $n$'th band has dispersion
\begin{align}
\label{eq:bandshape}
	E_n(k_x, \phi) &=  \frac{-1}{C'_n} (-1)^q\left[  P(\phi)^2 - 2P(\phi)\cos(k_x q) \right] + E_n
\end{align}
where $E_n$ is an $n$-dependent shift independent of momenta. 
Thus each band has the same shape up to shift $E_n$ and scale $1/C'_n$, up to higher order corrections in the $E$-dependence of $C$.

Although the absolute position and scale of each band depends in a detailed way on the parameters, the \emph{ratio} of the bandwidth in the $k_x$ and $\phi$ directions is independent of $n$ and therefore easy to compute. 
The bandwidth in the $k_x$ direction is given by the total variation of the Eq.~\ref{eq:bandshape} at fixed $\phi$. 
\begin{align}
	\Delta_{n,k_x}(\phi) = 4 |P(\phi)| / C'_n
\end{align}
Maximizing over $\phi$ gives,
\begin{align}
	\Delta_{n, k_x} = 16 \left(A_J/2\right)^q / C'_n
\end{align}
where we have assumed $q$ even (the result differs by an unimportant factor of 2 for $q$ odd).
Similar elementary considerations yield the maximum variation in the $\phi$ direction at fixed $k_x$,
\begin{align}
	\Delta_{n,\phi} =16 [ (A_J/2)^q + 1/2] [ (A_J/2)^q] / C'_n
\end{align}
where again we have assumed $q$ is even. 
Taking the ratio of the bandwidths in the $k_x$ and $\phi$ directions,
\begin{align}
r = \frac{\Delta_{n,k_x}}{\Delta_{n,\phi}} &= \frac{1}{(A_J/2)^q + 1/2}
\label{Eq:DeltaE_xp}
\end{align}
which holds for all bands $n$. 

From Eq.~\ref{Eq:DeltaE_xp}, it follows that the incommensurate TFIM is localized at all energies in the incommensurate limit ($q \to \infty$) for $A_J>2$.
The total dispersion in the $k_x$ direction at fixed $\phi$ (summed across bands) is exponentially smaller than that in the $\phi$ direction at fixed $k_x$. 
Since the total 2D variation is upper bounded by a function that at most increases linearly with $q$ (as each band has at most $O(1)$ bandwidth), the variation in the $\phi$ direction (which is exponentially larger in $q$ than that in the $k_x$ direction) provides the entire contribution as $q \to \infty$. 
Thus, the total bandwidth in the $k_x$ direction goes to zero exponentially as $q \to \infty$, so that the 1D incommensurate TFIM for $A_J>2$ has a pure-point spectrum.

For $A_J < 2$, the ratio approaches 1 and the variation of the bands is the same in both the $k_x$ and $\phi$ directions. 
In previously studied models, the 1D spectrum is critical whenever the ratio remains of  $O(1)$ in the incommensurate limit \cite{Han:1994cs}.
We conjecture and have confirmed numerically that this holds for the TFIM as well at $J=0$ (ie. that all states are critical at all energies). 
The coincidence of criticality and the order one ratio of bandwidths has not been proven mathematically in any model as far as we are aware.


\section{Evaluation of product of periodic couplings}
\label{app:EvalPk2}

In this appendix, we evaluate the expression:
\begin{align}
P(\phi) \equiv \prod_{j=0}^{q-1} ( J+A_J \cos(2\pi p (j+1/2)/q + \phi))
\end{align}
The derivation is identical to that in Appendix A of Ref.~\cite{Han:1994cs}; we include it here for completeness. First, $P(\phi)$ can be re-written as:
\begin{align}
P(\phi) = \prod_{j=0}^{q-1} \left[J + \frac{A_J}{2}\right.&\left(e^{i2\pi p(j+1/2)/q + i\phi}+ \right.\nonumber \\ 
&\left.\left.e^{-i2\pi p(j+1/2)/q - i\phi}\right)\right] \label{Eq:Pk2exp}
\end{align}
The expression is periodic in $\phi$ with period $2\pi/q$ as the addition of $2\pi/q$ to $\phi$ yields a permutation of the terms in the product. Thus, the only terms that survive the product in Eq.~\eqref{Eq:Pk2exp} contain $q$ factors of $e^{i\phi}$ or $q$ factors of $e^{-i\phi}$ or equal numbers of factors of $e^{i\phi}$ and $e^{-i\phi}$. The first two cases give:
\begin{align}
\label{Eq:Pk2_k2depterms}
\left(\frac{A_J}{2}\right)^q (-1)^{pq} (2\cos(q\phi))
\end{align}
The $\phi$ independent term can be obtained by evaluating $P(\phi)$ minus the term above at $\phi=0$:
\begin{align*}
Q = \prod_{j=0}^{q-1} \left[ J + A_J \cos\left(\frac{2\pi p(j+1/2)}{q} \right)\right] - 2(-1)^{pq} \left(\frac{A_J}{2}\right)^q
\end{align*}
Rearranging:
\begin{align*}
\frac{Q}{(A_J/2)^q} + 2 (-1)^{pq} = \prod_{j=0}^{q-1} \left[\frac{2 J}{A_J} + 2\cos\left(\frac{2\pi p}{q} \left(j+\frac{1}{2}\right)\right) \right]
\label{Eq:Qrearrange}
\end{align*}
The RHS is a polynomial in $ J/A_J$ with zeros at:
\begin{align}
\frac{J}{A_J} = -\cos\left(\frac{2\pi p}{q} \left(j+\frac{1}{2}\right)\right)
\end{align} for $j=0,\ldots, q-1$. Using the definition of the Chebyshev polynomial $T_q$ of order $q$:
\begin{align}
T_q(x) \equiv \cos(q \arccos(x)),
\end{align}
we see that $T_q( J/A_J) = (-1)^{pq+1}$ is a polynomial of $ J/A_J$ with the same zeros as the RHS of Eq.~\eqref{Eq:Qrearrange}. Thus, the two must be proportional. Comparing the coefficient of $( J/A_J )^{q}$ in the two expressions, we obtain:
\begin{align}
\frac{Q}{(A_J/2)^q} + 2 (-1)^{pq} &= 2[ T_q( J/A_J) - (-1)^{pq+1} ] \\
\Rightarrow Q &= 2 \left(\frac{A_J}{2}\right)^q T_q( J/A_J)
\end{align}
Combining this expression with Eq.~\eqref{Eq:Pk2_k2depterms} gives:
\begin{align}
P(\phi) = 2 \left(\frac{A_J}{2}\right)^q \left( T_q( J/A_J) + (-1)^{pq} \cos(q\phi) \right)
\label{Eq:Pk2FinalExpApp}
\end{align}
Using $(-1)^{pq} = (-1)^{p+q+1}$ when $p$ and $q$ are relatively co-prime, we obtain Eq.~\eqref{Eq:Pk2FinalExpMainText}.

\section{Ground state symmetry-breaking from the commensurate analysis}
\label{app:phaseboundaries}
In Appendix~\ref{app:EvalPk2}, we evaluated $P(\phi)=\prod_{m=0}^{q-1} J_{m+1/2}$ when the exchange coupling is commensurate with the underlying lattice at wavenumber $Q = 2\pi p/q$. 
In order to approach the incommensurate limit, we take $p,q \to \infty$ in such a way that the wavelength $2\pi/Q$ approaches an irrational number (in units of the underlying lattice constant $a=1$).
As $P(\phi)$ is independent of of the ratio $p/q$, the limiting expression is independent of the value of the irrational wavelength $2\pi/Q$. 
From the discussion in Sec.~\ref{sec:GroundStateSymmetryBreaking}, the behavior of $P(\phi)$ controls the ground state phase diagram: if $P(\phi)$ increases (decreases) exponentially with $q$, then the system is in the ferromagnetic (paramagnetic) phase.

From the explicit expressions:
\begin{align}
T_q(x) = \left\{ \begin{array}{cc}
\cos(q\arccos(x)) & |x|<1 \\
\frac{(x - \sqrt{x^2-1})^q + (x+\sqrt{x^2-1})^q}{2} & |x|>1,
\end{array}
\right.
\end{align}
the asymptotic forms of $P(\phi)$ at large $q$ follow:
\begin{align}
P(\phi) \sim \left\{ \begin{array}{cc}
				\left(\frac{A_J}{2}\right)^q &  J<A_J \\
                \left(\frac{ J + \sqrt{ J^2 - A_J^2}}{2}\right)^q & J>A_J
                \end{array}
                \right.
    \end{align}
    Note that $h=1$ in the expression above. Reinstating the $h$-dependence, we obtain the boundary between FM and PM ground states to be at $A_J/h=2$ for $0\leq  J/h \leq 2$ and at $ J/h + \sqrt{ (J/h)^2 - (A_J/h)^2} = 2$ for $A_J/h<2$. Simplifying the latter expression, the critical value of the exchange coupling is $ (J/h)_c = 1 + (A_J/2h)^2$ for $A_J/h\leq 2$.
    This is in perfect agreement with Eq.~\eqref{eq:gsphaseboundary}.

\end{document}